\documentclass{aa}  

\usepackage{graphicx}
\usepackage{txfonts}
\usepackage{soul}
\begin{document}

   \title{MN Draconis - peculiar, active dwarf nova in the period gap}

   \author{K. B\k{a}kowska,
          \inst{1},\inst{2}
          A. Olech,
          \inst{1}
          R. Pospieszy\'{n}ski,
          \inst{3} 
		 E.\'{S}wierczy\'{n}ski,
		 \inst{4} 
         F. Martinelli,
         \inst{5} 
		 A. Rutkowski,
		 \inst{6}
		 R. Koff,
		 \inst{7}
		 K. Drozd,
		 \inst{1}
         M. Butkiewicz-B\k{a}k 
          \inst{8}
          \and
         P. Kankiewicz\inst{9}
          }

   \institute{ Nicolaus Copernicus Astronomical Center,
				Polish Academy of Sciences,
				ul. Bartycka 18, 00-716 Warszawa, Poland\\
             \email{bakowska@camk.edu.pl}
         \and
             Fulbright Visiting Scholar, The Ohio State University, Dept. of Astronomy, 140 W. 18th Ave, Columbus, OH 43210, USA
		\and
		Comets and Meteors Workshop, ul. Bartycka 18, 00-716 Warszawa, Poland
		\and
		Centre for Astronomy, Faculty of Physics, Astronomy and Informatics, Nicolaus Copernicus University, Grudzi\k{a}dzka 5, 87-100 Toru\'{n}, Poland
		\and		
		Lajatico Astronomical Centre, Loc i Fornelli No. 9, Orciatico Lajatico, Pisa, Italy
		\and
Mt. Suhora Observatory, Pedagogical University, ul. Podchor\k{a}\.{z}ych 2, 30-084 Krak\'{o}w, Poland
		\and
	Center for Backyard Astrophysics, Antelope Hills Observatory, 980 Antelope Drive West, Bennett, CO 80102, USA
		\and
		Astronomical Observatory Institute, Faculty of Physics, A. Mickiewicz University, ul. S{\l}oneczna 36, 60-286 Pozna\'{n}, Poland
		\and
		Institute of Physics, Astrophysics Division, Jan Kochanowski University, \'{S}wi\k{e}tokrzyska 15, 25-406 Kielce, Poland
             }

\titlerunning{MN Draconis - a dwarf nova in the period gap}
\authorrunning{K. Bakowska et al.}

   \date{Received; accepted}
 
  \abstract
   {We present results of an extensive world-wide observing campaign of MN Draconis.}
   {MN Draconis is a poorly known active dwarf nova in the period gap and is one of the only two known cases of period gap SU UMa objects showing the negative superhumps. Photometric behaviour of MN Draconis poses a challenge for existing models of the superhump and superoutburst mechanisms. Therefore, thorough investigation of peculiar systems, such as MN Draconis, is crucial for our understanding of evolution of the close binary stars.}
   {To measure fundamental parameters of the system, we collected photometric data in October 2009, June-September 2013 and June-December 2015. Analysis of the light curves, $O-C$ diagrams and power spectra was carried out.}
   {During our three observational seasons we detected four superoutburts and several normal outbursts. Based on the two consecutive superoutbursts detected in 2015, the supercycle length was derived $P_{sc}=74 \pm 0.5$ days and it has been increasing with a rate of $\textit{\.{P}}=3.3 \times 10^{-3}$ during last twelve years. Based on the positive and negative superhumps we calculated the period excess $\varepsilon = 5.6\% \pm 0.1\%$, the period deficit $\varepsilon_{-} = 2.5\% \pm 0.6\%$, and in result, the orbital period $P_{orb}=0.0994(1)$ days ($143.126 \pm 0.144$ min). We updated the basic light curve parameters of MN Draconis. }
   {MN Draconis is the first discovered SU UMa system in the period gap with increasing supercycle length.}


  \maketitle

%

\section{Introduction}

Cataclysmic variables (CVs) are interacting binary systems containing a white dwarf primary and a low mass secondary. In these systems matter is transferred from the Roche lobe-filling secondary to the white dwarf. The accretion flows onto the poles of the primary or via an accretion disk depending on the strength of the magnetic field of the white dwarf. The region where the gas hits the edge of the disk is known as the hot spot. General reviews of CVs are presented by \cite{Warner1995} and \cite{Hellier2001}.

Dwarf novae (DN) are a subclass of non-magnetic CVs. In the light curves of these objects one can observe characteristic sudden increase in brightness known as outbursts. The thermal-viscous disk-instability model (DIM) explains outbursts as an abrupt increase in mass accretion through the disk, and in result the accretion disk brightens by factor 10-100 for a couple of days. In terms of the DIM, the mechanism responsible for the sudden increase in the mass transfer through the disk is a bimodal behaviour of the viscosity as a function of the surface density (see reviews by \citealt{Smak1984, Cannizzo1993, Lasota2001}).

Among DN there is a group known as SU UMa-type stars which are characterized by short orbital periods ($P_{orb} < 2.5 $ h). Moreover, SU UMa variables manifest not only outbursts but also about one magnitude brighter and less frequent superoutbursts. The time span between two successive superoutbursts is called the supercycle and on the basis of its duration further division of SU UMa class was made. ER UMa stars characterize by extremely short supercycles, which last several dozen of days, depending on the star. They are often referred to as active DN or erupting DN due to their nearly continuously outbursting behaviour. WZ Sge stars are the objects with the longest supercycle and in their light curves subsequent superoutbursts are commonly detected every $\sim$ 10 years. The nature of the length of the supercycle is still not fully understood. The situation is more complicated due to presence of quasi-periodic oscillations called superhumps in light curves of SU UMa systems during superoutbursts. Stars with positive superhumps, simply known as superhumps, exhibit oscillations with periods a few percent longer than their orbital periods. The interpretation of positive superhumps is that they arise from an apsidal advance (commonly referred as 'precession') of the eccentric accretion disk \citep{Vogt1982}. While outburst mechanism is understood by considering the thermal instability, the superoutbursts and superhumps are explained in terms of the tidal instability discovered and thoroughly investigated, i.e. \cite{Whitehurst1988, Hirose1990, Lubow1991}. In this scenario the accretion disk becomes distorted to eccentric shape. The occurrence of superhumps is caused by the periodic tidal stressing of the eccentric disk by the orbiting secondary. The tidal instability is due to the 3:1 resonance between the fluid flow in the disk and the orbital motion of the donor. At that moment the disk is large enough to reach the 3:1 radius ($r \sim 0.47a$, whereas $a$ is the binary separation), and the resonance sets in. This condition is only met for the mass ratio $q<0.25$, and hence only for CVs with low mass secondary components. The thermal-tidal instability model (TTI), proposed by \cite{Osaki1989}, explains the superoutburst cycle observed in SU UMa-type stars as caused by the thermal and tidal instabilities within the accretion disk. However, many observed features in SU UMa light curves pose serious challenge to the TTI model, i.e. the phenomenon of superoutbursts and superhumps in systems such as CzeV404 \citep{Bakowska2014} with the mass ratio $q$ higher than $0.25$. The idea of the enhanced mass transfer model (EMT) as a mechanism responsible for superoutbursts was first proposed by \cite{Vogt1983} and further developed by \cite{Osaki1985}. In this scenario irradiation heating of the secondary causes enhanced mass transfer, and therefore the superoutbursts in SU UMa-type systems. Despite the fact that \cite{Osaki1996} abandoned the concept of EMT, the model has been developed. Major weaknesses of the TTI model and its discrepancies with observations were presented, i.e. \cite{Smak1991, Smak1996, Smak2000, Smak2008, Bakowska2014b}.

In some cases of SU UMa systems we observe negative superhumps. These oscillations have periods which are slightly shorter than orbital periods. They occur as a result of nodal retrograde wobble (denoted also as 'regression') of the tilted disk. However, no theory of negative superhumps has been well established so far. \cite{Barrett1988} proposed that the source of negative superhumps is the varying energy of the gas stream as it impacts on face of the disk. \cite{Patterson1997} suggested that the source of the negative superhumps is gravitational energy. If the disk is tilted then the gas stream can easily flow over the top and hit the inner disk. \cite{Wood2007} claim that the negative superhump modulation arises from the hot spot transiting across each face of the tilted disk. Recently, \cite{Thomas2015} presented the SPH simulations showing that the magnetic field on the primary star  enables the disk to tilt causing retrograde precession of the disk and leads to the emergence of negative superhumps. However, according to \cite{Thomas2015}, the superhump period is determined by the presence of the secondary component not by the magnetic field of the white dwarf.   

Another issue in the evolution of CVs is the detection of active DN in the 2-3 hour orbital period range, known as the orbital period gap. The magnetic braking and the gravitational radiation are the two mechanisms which are thought to drive the orbital angular momentum loss. Both of them are greatly reduced in the period gap region, and hence the number of CVs found is very low (see \citealt{Gansicke2009}). According to the TTI model systems with orbital periods ($\approx 2.5$ h) should be characterized by low activity. Nonetheless, recent discoveries of active DN in the period gap region pose problem for existing theory, e.g. CzeV404 \citep{Bakowska2014}.

MN Draconis (MN Dra) is another example of an active SU UMa-type DN in the period gap.
The first intensive photometric observation campaign was made by \cite{Nogami2003}. They detected three superoutbursts in October 2002, December 2002, and February 2003, respectively. Moreover, \cite{Nogami2003} derived the supercycle period of $P_{sc} \sim 60$ days, one of the shortest among the values known in SU UMa stars. Also, they measured the superhump periods $0.104885(93)$ days and $0.10623(16)$ days during the first and the second superoutbursts, respectively. Additionally, \cite{Nogami2003} estimated the rate of changes of the superhump period which was increasing during the first superoutburst and decreasing during the second superoutburst. It is also worth to mention their detection of $0.10424(3)$ days periodicity in quiescence. They concluded that MN Dra is related to ER UMa objects in terms of the supercycle length and the normal cycle length.   

The next observing campaign dedicated to MN Dra was organized by the Russian team. \cite{Pavlenko2010} observed MN Dra during $18$ nights between May and June 2009. They reported one superoutburst in May 2009 with the corresponding superhump period $0.105416$ days decreasing with the rate of $-24.5 \times 10^{-5}$. In quiescence, the object manifested signal of $0.09598$ days. From the harmonic analysis-of-variance periodograms, \cite{Pavlenko2010} estimated the orbital period of $0.0998(2)$ days. Later, \cite{Samsonov2010} detected two superoutbursts and five outbursts in MN Dra during $77$ nights of observations in the August - November 2009 season. They noticed a manifestation of the positive superhumps with a period of $0.105416$ days during superoutbursts with a decreasing rate of $-(3-8) \times 10^{-4}$, and the negative superhumps with a period of $0.095952$ days during quiescence and normal outbursts.

The latest information about MN Dra was presented by \cite{Kato2014b}. They reported two superoutbursts of MN Dra in July - August 2012 and in November 2013. In both cases the superhump period was obtained ($0.105299(61)$ days and $0.105040(66)$ days, respectively). Also, based on the orbital period estimation given by \cite{Pavlenko2010}, \cite{Kato2014b} calculated mass ratio ($q=0.327$ and $q=0.258$) of MN Dra during its July-August 2012 and November 2013 superoutbursts, respectively. 

MN Dra is a very useful target for exploring the aforementioned (and related)
issues. Not only is MN Dra one of the rare SU UMa systems in the period
gap, it is one of the few DN exhibiting both positive and negative superhumps.
We have therefore characterized the occurrence of superhumps in superoutburst,
in outburst, and in quiescence, along with their rates of change. We have also
refined the supercycle length (and possible changes in superoutburst occurrence
rate). The structure of this paper is following: Section \ref{Obs and Data Red.} presents details about observations, data reduction and photometry analysis. Section \ref{Light Curves} contains information about photometric behaviour of the system and thorough study of positive and negative superhumps. Discussion is shown in Section \ref{Discussion}. We present summary of our campaign dedicated to MN Dra in Section \ref{Conclusions}.


\section{Observations and Data Reduction}
\label{Obs and Data Red.}

Observational data of MN Dra presented here were collected throughout three observation campaigns. The first data were obtained  during 6 nights from 2009 October 12 to October 22, in the Skinakas Observatory, Greece. At that time, the star was in a quiescent state. The second campaign was conducted during 15 nights from 2013 July 09 to September 10. On that occasion, data were gathered in Poland, at the Borowiec station of the Pozna\'{n} Astronomical Observatory, at J. Kochanowski University in Kielce, and at the Ostrowik station of the Warsaw University Observatory. During the 2013 campaign we detected two superoutbursts of MN Dra. The longest campaign was organized in 2015, from June 4 to December 18. Data covering 84 nights of observations were collected at the Borowiec Station, in Pisa in Italy, at Antelope Hills Observatory and at the MDM Observatory in the USA. Once again, two superoutbursts in MN Dra were observed.

In Table~\ref{tab:MNDra_log}, we present the journal of CCD observations of MN Dra. In total, the star was observed 374.37 hours and 8717 useful exposures were obtained. To summarize, eight telescopes with diameters ranging from 0.25 to 1.30-meter, were used to collect data during the 2009, 2013 and 2015 campaigns.  

\begin{table*}
	\centering
	\caption{The journal of our CCD observations of MN Dra in 2009 - 2015.}
	\label{tab:MNDra_log}
	\begin{tabular}{llrrrl} 
		\hline
		\noalign{\smallskip}
		Observatory  & Telescope & No. of & Total time & No. of & Observer\\
		(Country)& (meter) & nights & [h] & frames\\
		\noalign{\smallskip}
		\hline
		\noalign{\smallskip}
		Toru\'{n} Centre of Astronomy  & 0.60  & 31 & 100.94 & 1881 & E. \'Swierczy\'nski, K. Drozd\\
		(Poland) &&&&& \\
		\noalign{\smallskip}		
		Pozna\'{n} Astronomical Obs.  & 0.40  & 24 & 63.89 & 1137 & R. Pospieszy\'nski, K. B\k{a}kowska,\\
		(Poland) &&&&& M. Butkiewicz-B\k{a}k\\
	    \noalign{\smallskip}		
		Warsaw University Obs.  & 0.60  & 10 & 63.70 & 2272 & K. B\k{a}kowska\\
		(Poland) &&&&& \\
	    \noalign{\smallskip}		
		Lajatico Astronomical Obs.  & 0.50 & 19 & 59.99  & 911 &  F. Martinelli\\
		(Italy) &&&&& \\
	    \noalign{\smallskip}		
		MDM Obs. & 1.30  & 8 & 23.58  & 1202 &  K. B\k{a}kowska\\
		 (USA) &&&&& \\
	    \noalign{\smallskip}		
		Skinakas Obs.  & 1.30  & 6 & 21.40  & 371 &  A. Rutkowski\\
		(Greece) &&&&& \\
		 \noalign{\smallskip}		
		Antelope Hills Obs. & 0.25 & 5 & 37.01 & 837 & R. Koff\\
		 (USA) &&&&& \\
		 \noalign{\smallskip}		
		J. Kochanowski University& 0.35 & 3 & 3.86  & 106 &  P. Kankiewicz\\
		 (Poland) &&&&& \\
		\noalign{\smallskip}		
		\hline
	\end{tabular}
\end{table*}

\begin{figure}
	\includegraphics[width=\columnwidth]{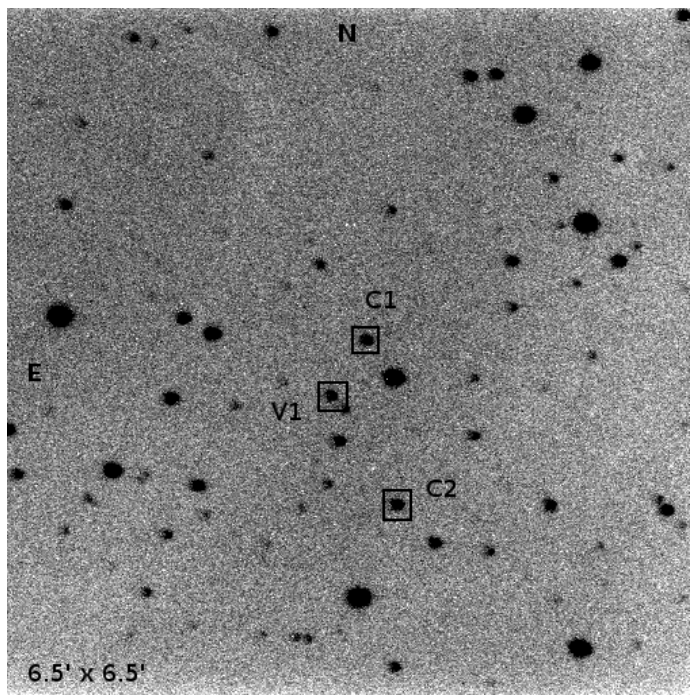}
    \caption{Finding chart with a position of MN Dra marked as V1 and two comparison stars C1 and C2. The field of view is about 6.5'$\times$6.5'.}
    \label{fig:SkyChart}
\end{figure}

All data of MN Dra were gathered in a clear filter ('white light'). This approach allowed us to obtain reliable photometry around 18-20 mag when the object was in quiescence. Bias, dark and flat-field correction of raw files was done in a standard way. For data reduction we used the IRAF package. The profile (point spread function) photometry was obtained with the DAOPHOTII package \citep{Stetson1987}. Relative unfiltered magnitudes of MN Dra were derived from the difference between the magnitude of the star and the mean magnitude of two comparison stars marked as C1 and C2 on Fig.~\ref{fig:SkyChart}, where the map of the observed region is shown. The equatorial coordinates of comparison stars C1 (RA=$20^{h}23^{m}35^{s}.33$,
Dec=$+64^{o}36'56".3$, $15.800$ mag in $R$ filter) and C2 (RA=$20^{h}23^{m}32^{s}.59$,
Dec=$+64^{o}35'29".1$, $15.800$ mag in $R$ filter) were taken from 2MASS All-Sky Catalog \citep{Cutri2003}. Instrumental magnitudes were transformed into the R pass-band which is very close to observations made in 'white light'.

\section{Light Curves}
\label{Light Curves}

\subsection{Global photometric behaviour}

\begin{figure*}
	\vspace{0.75cm}
	\centering
	\includegraphics[scale=0.7]{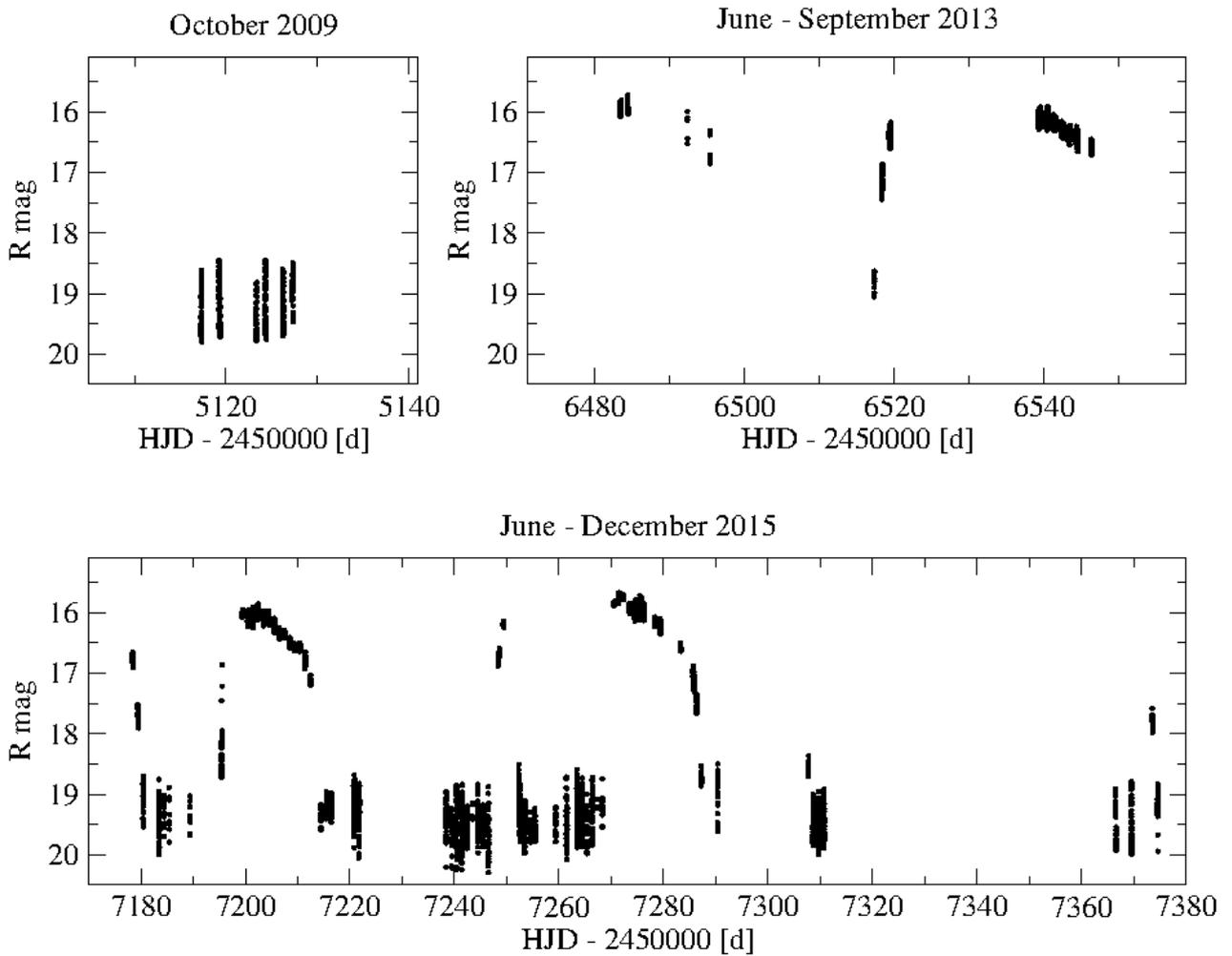}
    \caption{The global photometric behaviour of MN Dra.}
    \label{fig:GlobalLightcurve}
\end{figure*}

For convenience, we use only a day number HJD-2450000 [d] to refer to our observations. Fig.~\ref{fig:GlobalLightcurve} shows the light curves of MN Dra during our three campaigns. 

At the time of observations in 2009 (top left panel of Fig.~\ref{fig:GlobalLightcurve}) MN Dra was at quiescence but we detected negative superhumps (more in Sections \ref{sub:ANOVA} and \ref{sub:OC}). 

During the 2013 campaigns (Fig.~\ref{fig:GlobalLightcurve}, top right panel) we observed two superoutbursts interspersed with one normal outburst. Due to unfavourable weather conditions, we obtained only 4 nights of data covering the first July 2013 superoutburst HJD 6482, 6483, 6492 and 6495. However, clear superhumps were observed on HJD 6483. Later on, MN Dra was caught on the rise to normal outburst during the HJD 6517, 6518 and 6519 nights. The last data from the 2013 campaign containing 7 nights (between HJD 6539 and HJD 6546) showed the object in the September 2013 superoutburst. Also, this time the superhumps were clearly visible in each run. 

During the 2015 campaign (bottom panel), we gathered 14 nights (HJD 7199 - 7212) of the June - July superoutburst, and 13 nights (between HJD 7268 and HJD 7286) of the September superoutburst. Additionally, we detected three normal outbursts that occurred in June, August and December 2015.

During maximum brightness MN Dra reached $R \approx 15.7$ mag, and  after its superoutbursts the star faded to $R \approx 18.6$ mag, resulting with the average amplitude of superoutburst $A_s \approx 2.9$ mag. In quiescence, brightness of MN Dra was varying between $R \approx 18.4$ and $R \approx 20.3$ mag. Because of diverse weather during several normal outbursts, any estimations of outburst amplitudes and their maximum and minimum levels were impossible to derive.

\subsection{Supercycle length}
\label{Supercycle_Length}

Until today, the supercycle length of MN Dra has been measured twice. The first one, $P_{sc1} \sim 60$ days, was derived by \cite{Nogami2003}, with no given uncertainty. 

The second estimation was made by \cite{Samsonov2010}. They reported the length of the supercycle of $\sim 30$ days. Nonetheless, based on the visual inspection of the light curve presented in Fig. 1 \citep{Samsonov2010} we consider this value as completely incorrect. Hence, we re-analysed the supercycle length and obtained a rough estimation of $P_{sc2} \sim 67.5 \pm 2.5$ days. 

We derived $P_{sc3}$ from our data from 2015, using the ANOVA code of \citep{Schwarzenberg1996}. The most prominent peak of the power spectrum was found at the frequency $0.0135(1)$ [c/d] and this corresponds to $P_{sc3}=74\pm0.5$ days.

Due to the fact that our estimate of $P_{sc3}$ is based on only two superoutbursts observed in 2015, we decided to investigate our result more precisely. First, we looked for the amateur data of MN Dra in superoutbursts in the databases such as the AAVSO covering the $2014 - 2016$ observational seasons. Unfortunately, this object was not observed by amateurs. 
Without any additional observations, we decided to calculate the supercycle length based on our data set covering all four superoutbursts which we detected in the 2013 and 2015 observational campaigns. We obtained the supercycle period of $\sim 72$ days. Not only does this crude estimation confirm that the supercycle length of MN Dra is constantly increasing but also shows that the timespan between two successive superoutbursts is currently longer than $72$ days. Including our data set, the corresponding increasing rate of the supercycle length is $\textit{\.{P}}= 3.3 \times 10^{-3}$. Therefore, we consider $P_{sc3}=74\pm0.5$ days as reasonable and the most up-to-date value of the supercycle length of MN Dra. 

It is worth noting that in the light curves covering subsequent superoutbursts presented by \cite{Nogami2003} and \cite{Samsonov2010} are significant gaps, in particular the lack of the beginnings or endings of the superoutbursts. That is why, we made one more test for the values of the supercycle length. This time we calculated the time span between the last nights of quiescence between subsequent superoutbursts. Also, we derived the time span between the first nights of the quiescence after superoutbursts. Based on this analysis, the supercycle lengths were found
 $P_{sc1b}=62$ days for data set presented by \cite{Nogami2003} and  $P_{sc2b}=68$ days for light curves given by \cite{Samsonov2010}. With our data set, the rate of increase of supercycle length was $\textit{\.{P}}= 2.8 \times 10^{-3}$.

\begin{figure}
	\includegraphics[width=\columnwidth]{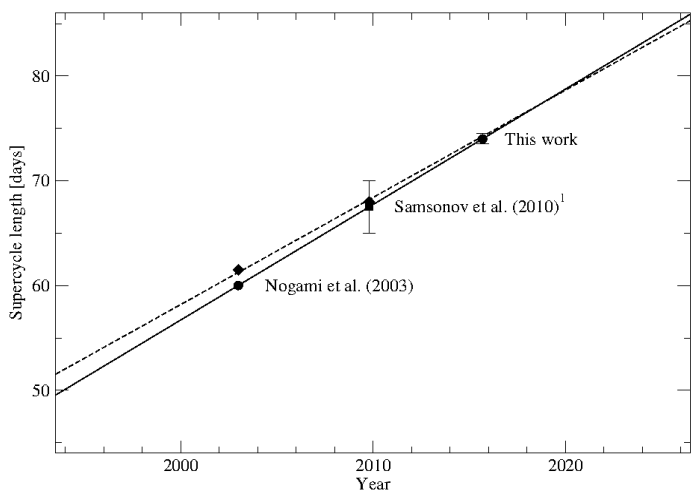}
    \caption{The increasing supercycle length of MN Dra during last twelve years. Lines correspond to the best fits to the data (details in text). Uncertainties are given when available. $^1$ This value of supercycle length was calculated by the authors of this work based on the light curves presented in \cite{Samsonov2010}.}
    \label{fig:SupercycleLength}
\end{figure}

In Fig.~\ref{fig:SupercycleLength} we present all available measurements of $P_{sc}$. The black line represents fit calculated based on $P_{sc1}$, $P_{sc2}$ and $P_{sc3}$ days. The dotted line corresponds to the fit obtained based on $P_{sc1b}$, $P_{sc2b}$ and $P_{sc3}$ days. In both cases, the immediate conclusion is that the supercycyle length of MN Dra has increased during last twelve years. The corresponding rate of the increase of the period is $\textit{\.{P}}= 2.8 - 3.3 \times 10^{-3}$.

\begin{figure}
	\includegraphics[width=\columnwidth]{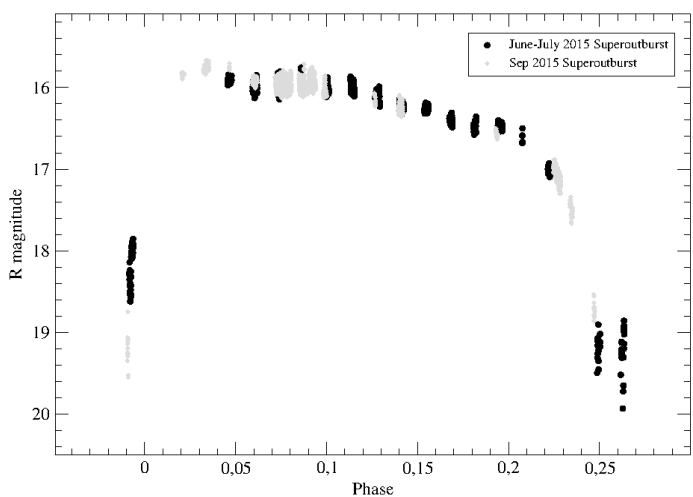}
    \caption{ The light curve of MN Dra during superoutbursts folded with $P_{sc}=74$ days. Black circles correspond to the June - July 2015 superoutburst and grey circles represent the September 2015 superoutburst.}
    \label{fig:FoldedLC}
\end{figure}

The $P_{sc3}$ was used to create the folded light curve of the June - July 2015 and the September 2015 superoutbursts in MN Dra. The result is presented in Fig.~\ref{fig:FoldedLC}.

\subsection{Superhumps}
\label{Superhumps}

\begin{figure*}
	\centering
	\includegraphics[scale=0.45]{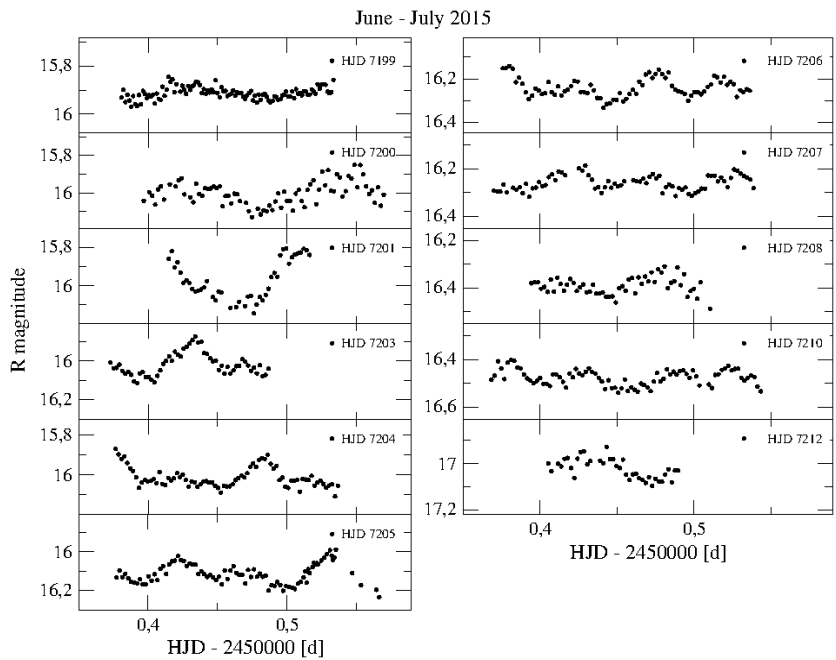}
    \caption{The light curves of MN Dra during the June - July 2015 superoutburst. A fraction of HJD is presented on the $x$-axis. HJD-2450000 [d] is given on the right side of each panel.}
    \label{fig:Superhumps_June2015}
\end{figure*}

\begin{figure*}
	\centering
	\includegraphics[scale=0.45]{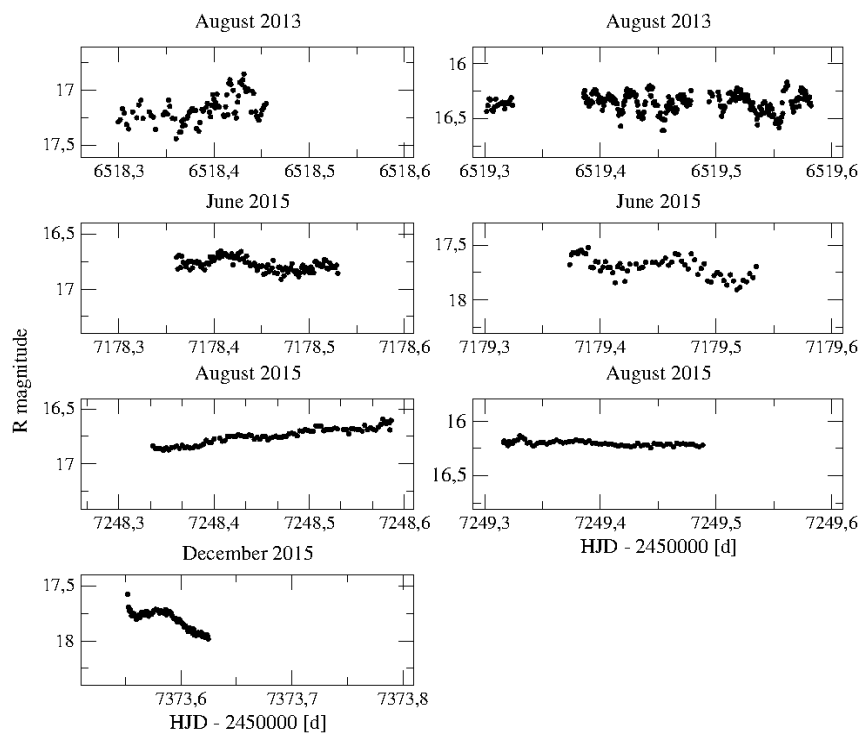}
    \caption{The light curves of MN Dra during its normal outbursts observed in August 2013, June 2015, August 2015, and December 2015.}
    \label{fig:Superhumps_NormalOutburst}
\end{figure*}

\begin{figure*}
	\centering
	\includegraphics[scale=0.45]{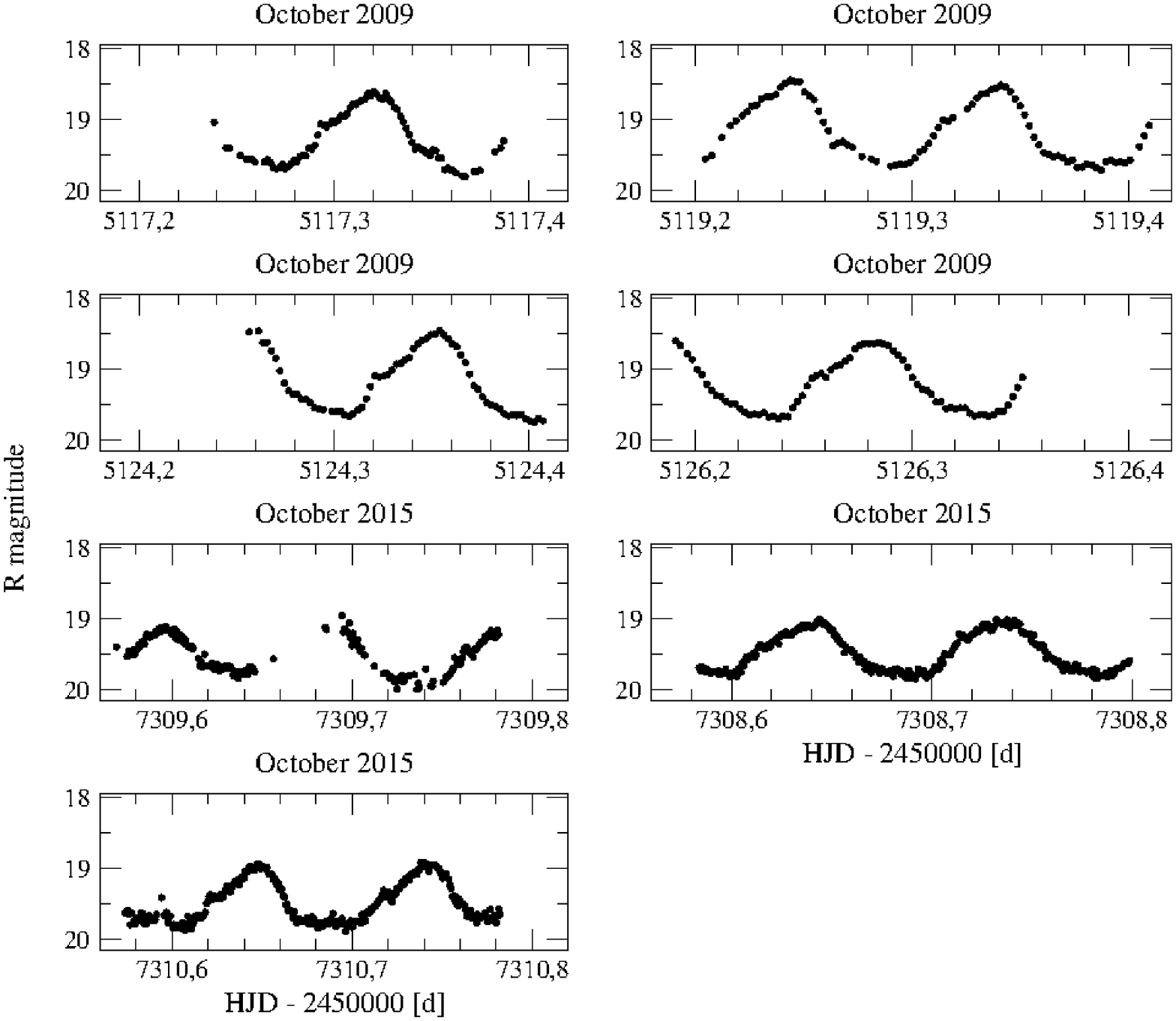}
   \caption{The light curves of MN Dra in quiescence. The October 2009 data set was obtained on 1.3-meter telescope in Greece. The October 2015 observations were gathered on the 1.3-meter telescope located in the U.S.A.}
   \label{fig:Superhumps_Quiescence}
\end{figure*}

In the light curves of MN Dra one can see clear superhumps. 
Examples of these characteristic 'tooth-shaped' oscillations are presented in Fig.~\ref{fig:Superhumps_June2015}, Fig.~\ref{fig:Superhumps_NormalOutburst} and 
Fig.~\ref{fig:Superhumps_Quiescence} during superoutburts, normal outbursts and in quiescence, respectively. 

We performed power spectra and the $O-C$ analysis for detected periodicities. 
The following Sections present results of this investigation.

\subsection{Period Analysis}
\label{sub:ANOVA}

We detrended the light curves of MN Dra by subtracting first or second order polynomials. Data from each observatory and from each night were converted one at a time. Then, we analysed them using the ANOVA statistics with one or two harmonic Fourier series \citep{Schwarzenberg1996}.    

Due to the different levels of quality of the collected data and various types of periodicities we investigated them separately. Our data were divided into six sets:
\begin{itemize}
\item the September 2013 superoutburst
\item the June - July 2015 superoutburst
\item the September 2015 superoutburst
\item normal outbursts
\item the October 2009 quiescence
\item the October 2015 quiescence
\end{itemize}

\subsubsection{Superoutbursts}

For the September 2013 and the June - July 2015 superoutbursts the most prominent peak was found at the frequency $f_{sh1}=9.510(7)$ [c/d] and $f_{sh2}=9.510(10)$ [c/d], respectively. This corresponds to a period $P_{sh1}=0.10515(8)$ days and $P_{sh2}=0.10515(11)$ days. In case of the September 2015 superoutburst the highest peak was detected at the frequency $f_{sh3}=9.300(90)$ [c/d] which determines the period of $P_{sh3}=0.10753(104)$ days. 

We adopt $P_{sh1}=P_{sh2}$ as the superhump period of MN Dra. Not only has it appeared in two superoutbursts with the exact same value but also the time coverage of observations collected during the September 2015 superoutburst was worse in comparison to the September 2013 and the June-July 2015 superoutbursts. 

The prewhitening procedure of the detrended light curves was performed to look for additional periodicities. We removed the modulation of the superhump period and then conducted the  ANOVA analysis. The resulting power spectra have intricate structures in all three cases of investigated superoutbursts. Despite the fact that we scrutinized several peaks in prewhitened spectra, we did not find any physical interpretation of their presence.

\begin{figure}
	\includegraphics[width=\columnwidth]{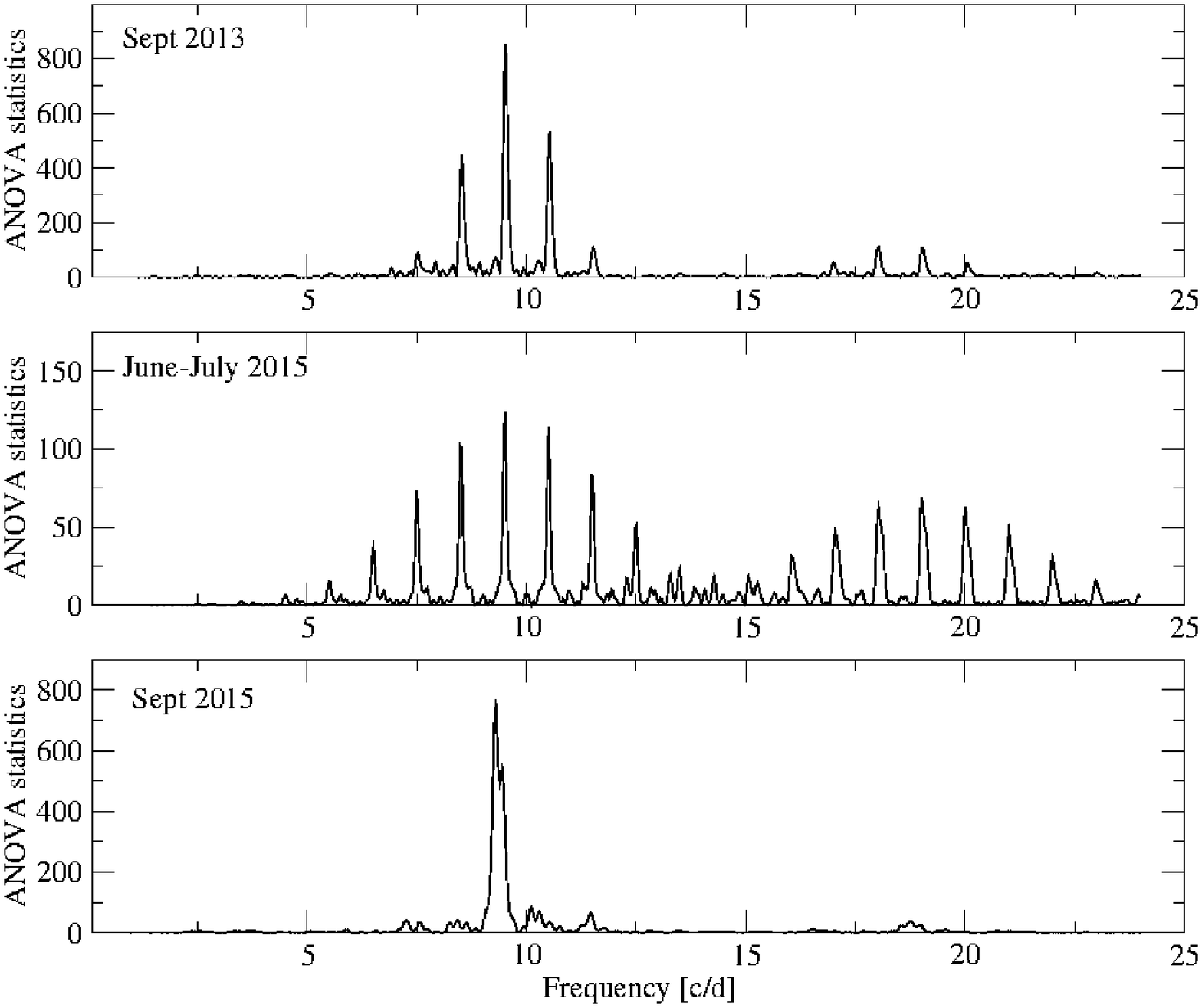}
    \caption{The ANOVA power spectra of the light curves of MN Dra during the September 2013, the June - July 2015 and the September 2015 superoutbursts.}
    \label{fig:MNDra_Anova_Perio_Superoutbursts}
\end{figure}

\begin{figure}
	\includegraphics[width=\columnwidth]{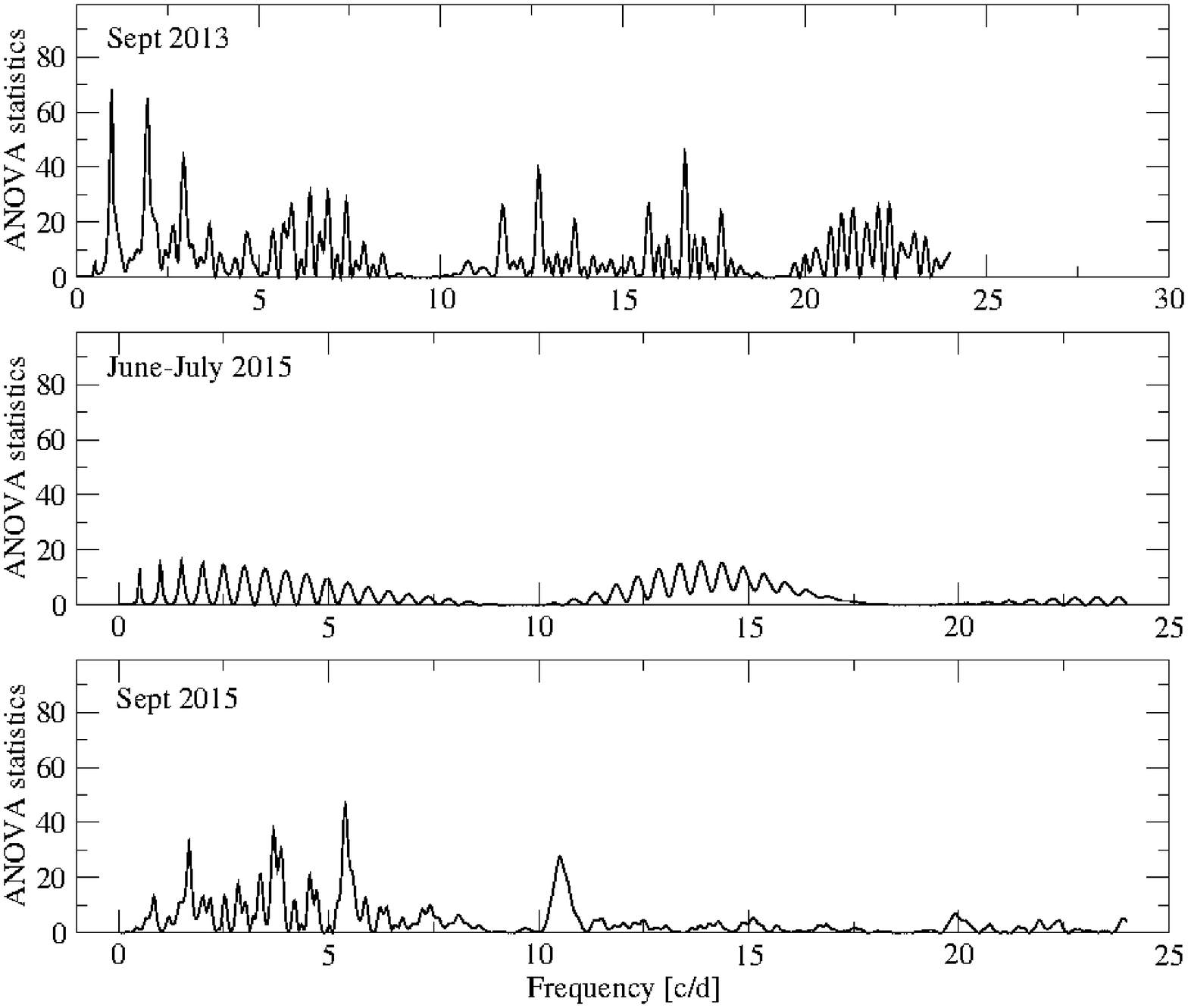}
    \caption{The ANOVA power spectra of the prewhitened light curves of MN Dra during the September 2013, the June - July 2015 and the September 2015 superoutbursts.}
    \label{fig:MNDra_Anova_Pre_Superoutbursts}
\end{figure}

Fig.~\ref{fig:MNDra_Anova_Perio_Superoutbursts} shows the power spectra for the light curves of the September 2013 (top panel), the June - July 2015 (middle panel) and the September 2015 (bottom panel) superoutbursts in MN Dra. Fig.~\ref{fig:MNDra_Anova_Pre_Superoutbursts} present periodograms for prewhitened light curves  of the September 2013 (top), the June - July 2015 (middle) and the September 2015 (bottom) superoutbursts in MN Dra.

\subsubsection{Normal outbursts}

\begin{figure}
	\centering
	\includegraphics[width=0.5\columnwidth]{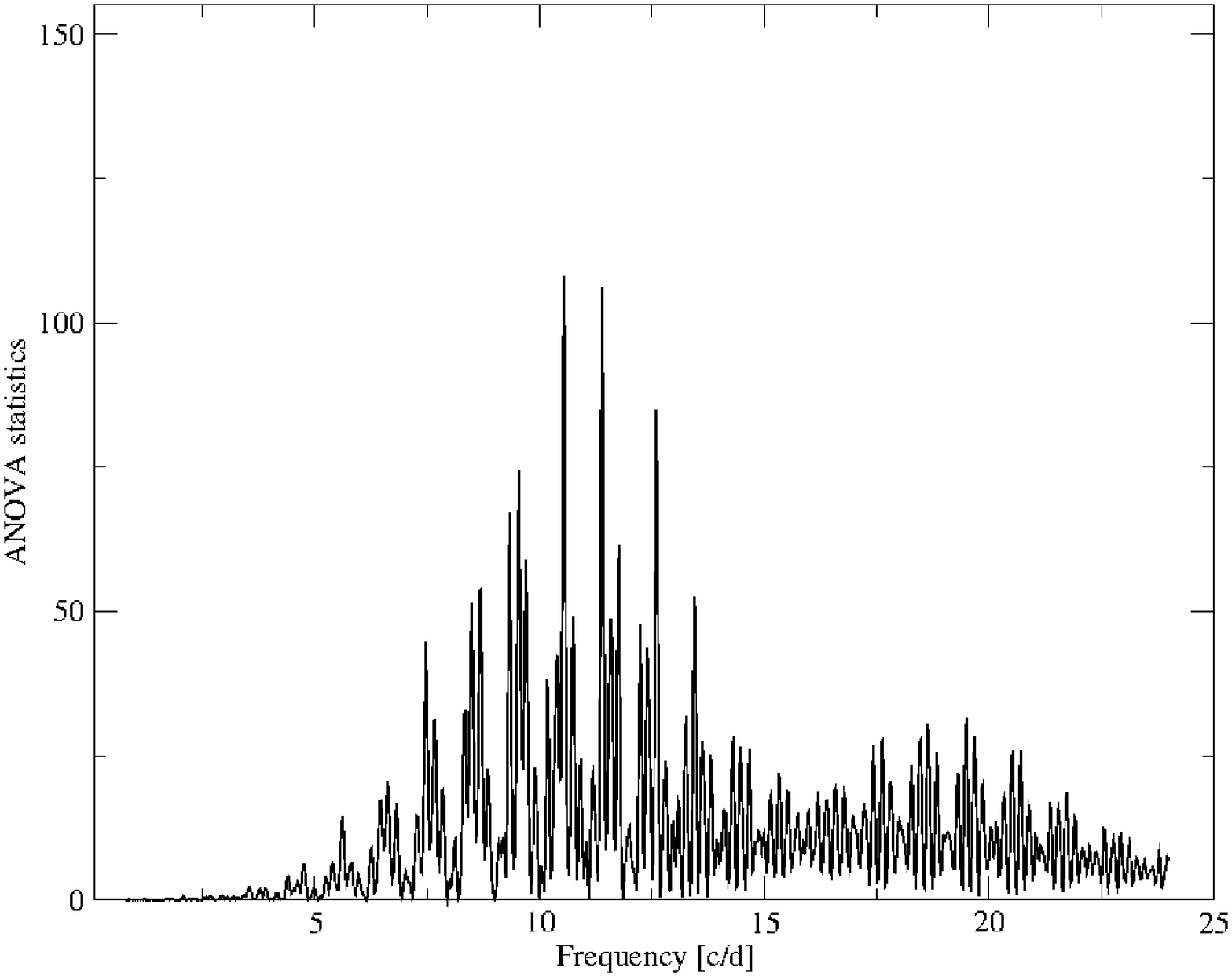}
    \caption{The ANOVA power spectra of the light curves of MN Dra during its normal outbursts.}
    \label{fig:MNDra_Anova_Perio_NO}
\end{figure}

\begin{figure}
	\centering
	\includegraphics[width=0.5\columnwidth]{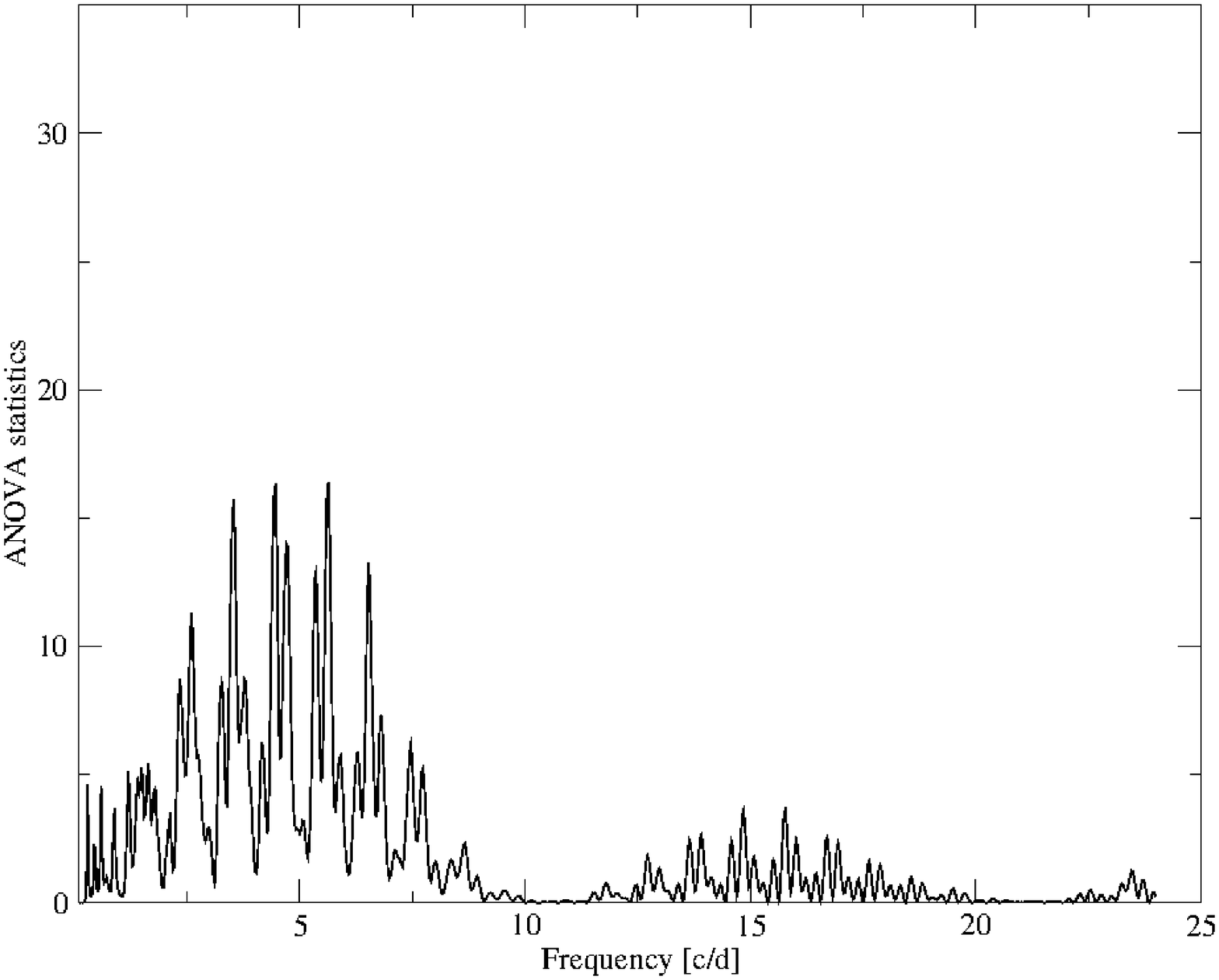}
    \caption{The ANOVA power spectra of the prewhitened light curves of MN Dra during its normal outbursts.}
    \label{fig:MNDra_Anova_Pre_NO}
\end{figure}

Due to overcast weather during every normal outburst we were not able to investigate them separately. Based on the all collected data, presented in Fig.~\ref{fig:Superhumps_NormalOutburst}, we performed the power spectra analysis. Fig.~\ref{fig:MNDra_Anova_Perio_NO} presents the resulting periodogram with the highest peaks located at frequencies $10.530(3)$ [c/d] and $11.400(5)$ [c/d], respectively. However, we can only interpret the first value $f_{nsh1}=10.530(3)$ [c/d] as the corresponding to the negative superhump period $P_{nsh1}=0.09497(3)$ days. Fig.~\ref{fig:MNDra_Anova_Pre_NO} shows the outcome of the prewhitening analysis. Again, we interpret the detected peaks as probably false signals.

\subsubsection{Quiescence}

The best quality data of MN Dra at minimum brightness were gathered on 1.3 meter telescopes located in Greece and in the U.S.A. (check Table \ref{tab:MNDra_log} and Fig. \ref{fig:Superhumps_Quiescence}). Hence, we performed two separate power spectra analysis for data sets covering light curves collected in October 2009 in Crete and in October 2015 in Arizona.

\begin{figure}
	\centering
	\includegraphics[width=\columnwidth]{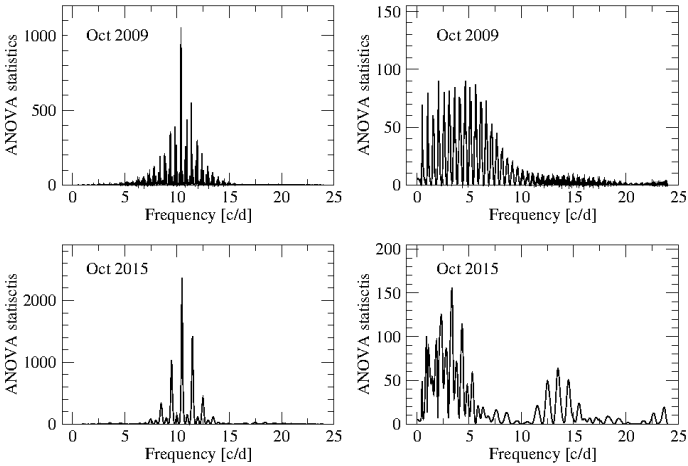}
    \caption{The ANOVA power spectra of the light curves (left panels) and prewhitened light curves (right panels) of MN Dra during in quiescence.}
    \label{fig:MNDra_Anova_Pre_Qui}
\end{figure}

Top left panel of Fig. \ref{fig:MNDra_Anova_Pre_Qui} shows the resulting power spectrum of the light curves from the October 2009 data set.  The most prominent frequency was detected at $f_{nsh2}=10.500(2)$ [c/d] and it is associated with the negative superhump period $P_{nsh2}=0.09524(2)$ days. The harmonic analysis-of-variance periodogram for prewhitened light curves of MN Dra collected in Greece is presented in the top right panel of Fig. \ref{fig:MNDra_Anova_Pre_Qui}. This forest of frequencies is caused by the changes in amplitude and the period of negative superhumps. The occurrence of negative humps is not a strictly periodic phenomenon, and hence it is impossible to subtract its value with only one analytical fit. As a result, we observe residua on the lower frequencies in the periodogram.

The result of the investigation of the October 2015 light curves based on the ANOVA statistics is displayed on the left bottom panel of Fig. \ref{fig:MNDra_Anova_Pre_Qui}. This time, the highest signal was found at $f_{nsh3}=10.470(6)$ [c/d] corresponding to $P_{nsh3}=0.09551(6)$ days. In the prewhitened spectra, presented on the right bottom panel of  Fig. \ref{fig:MNDra_Anova_Pre_Qui}, we detected the highest signal at $f=3.360(10)$ [c/d]. Despite our efforts, no physical interpretation was associated to this frequency.

In Table \ref{tab:MNDra_ANOVA} we present the detected periodicities in the light curves of MN Dra during its superoutbursts, normal outbursts and in quiescence.

\begin{table*}
	\centering
	\caption[The most prominent frequencies detected in the light curves of MN Draconis]{The most prominent frequencies detected in the light curves of MN Dra.}
	\label{tab:MNDra_ANOVA}
	\begin{tabular}{lll} 
		\hline
		\noalign{\smallskip}
		 & Frequencies [c/d] & Periods [days]\\
		 \noalign{\smallskip}
		\hline
		\noalign{\smallskip}
		September 2013 superoutburst  & $f_{sh1}=9.510(7)$ & $P_{sh1}=0.10515(8)$\\
		\noalign{\smallskip}
		June-July 2015 superoutburst  & $f_{sh2}=9.510(10)$ & $P_{sh2}=0.10515(11)$\\
		\noalign{\smallskip}		
		September 2015 superoutburst  & $f_{sh3}=9.300(90)$ & $P_{sh3}=0.10755(104)$\\
		\noalign{\smallskip}		
		normal outbursts  & $f_{nsh1}=10.530(3)$ & $P_{nsh1}=0.09497(3)$\\
		\noalign{\smallskip}		
		October 2009 quiescence  & $f_{nsh2}=10.500(2)$ & $P_{nsh2}=0.09524(2)$\\
		\noalign{\smallskip}		
		October 2015 quiescence  & $f_{nsh3}=10.470(6)$ & $P_{nsh3}=0.09551(6)$\\
		\noalign{\smallskip}			
		\hline
	\end{tabular}
\end{table*}

\subsection{The $O-C$ Diagram for Superhumps}
\label{sub:OC}

We observed positive and negative superhumps in the light curves of MN Dra.
To check the stability of the detected oscillations, the $O-C$ analysis for these periodicities was performed.

\subsubsection{Superoutbursts}

Here, we present results only for the September 2013 and the June - July 2015 superoutbursts. Due to unfavourable weather conditions, the September 2015 data were scattered and of a lower quality than the observations collected during the September 2013 and the June - July 2015 superoutbursts.

We identified 16 moments of maxima in the light curves of MN Dra during its September 2013 superoutburst. A linear fit allowed us to obtain the following ephemeris:

\begin{equation}
{\rm HJD_{\rm max}} = 2456539.3276(6) + 0.10496(2) \times E.
\label{eq:MNDra_OC_Max1}
\end{equation} 

Additionally, the second-order polynomial fit was calculated for the moments of maxima and the corresponding ephemeris was obtained:

\begin{equation}
{\rm HJD_{\rm max}} = 2456539.3266(7) + 0.10518(8) \times E - 5.6(1.9) \times 10^{-6} \times E^2.
\label{eq:MNDra_OCsq1}
\end{equation} 

In the data set of the June - July 2015 superoutburst, we determined 9 peaks of maxima. The following ephemeris was derived:

\begin{equation}
{\rm HJD_{\rm max}} = 2457199.4384(15) + 0.10512(3) \times E.
\label{eq:MNDra_OC_Max2}
\end{equation} 

Once again, we calculated the second-order polynomial fit for the moments of maxima and we obtained:

\begin{equation}
{\rm HJD_{\rm max}} = 2457199.4359(20) + 0.10529(9) \times E - 2.0(1.0) \times 10^{-6} \times E^2.
\label{eq:MNDra_OCsq2}
\end{equation}

Based on the derived ephemerides presented in Eq. \ref{eq:MNDra_OC_Max1} and \ref{eq:MNDra_OC_Max2}, the corresponding superhump periods were obtained: $P_{sh4}=0.10496(2)$ days ($151.1424\pm 0.03$ min) and $P_{sh5}=0.10512(3)$ days ($151.3728\pm 0.04$ min) for the September 2013 and the June - July 2015 superoutbursts, respectively.

Tables \ref{tab:OC_Sep2013} and \ref{tab:OC_Superoutburst_Jun2015} list cycle numbers $E$, times of detected maxima, errors, humps amplitudes and the $O-C$ values for the September 2013 and the June - July 2015 superoutbursts, respectively.  

\begin{figure}
	\centering
	\includegraphics[width=\columnwidth]{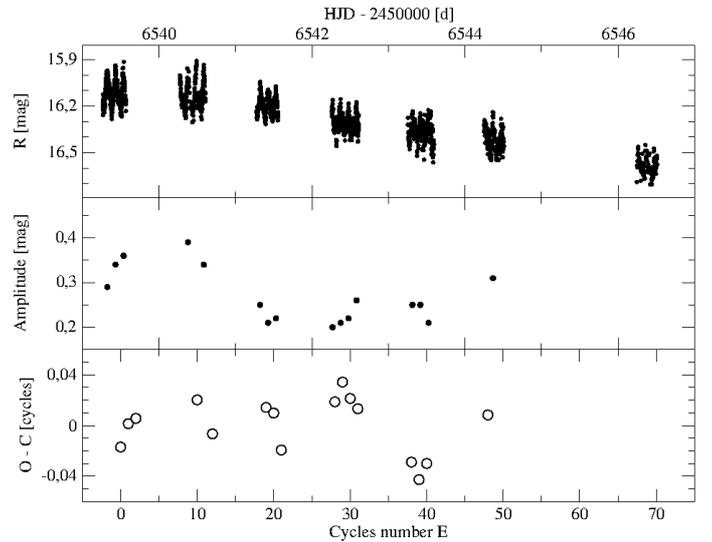}
    \caption{The light curves of MN Dra during the September 2013 superoutburst (top panel), the evolution of the amplitude of superhumps (middle panel) and the $O-C$ diagram for the superhumps maxima (bottom panel).}
    \label{fig:MNDra_OC_LC1_superout}
\end{figure}

\begin{figure}
	\centering
	\includegraphics[width=\columnwidth]{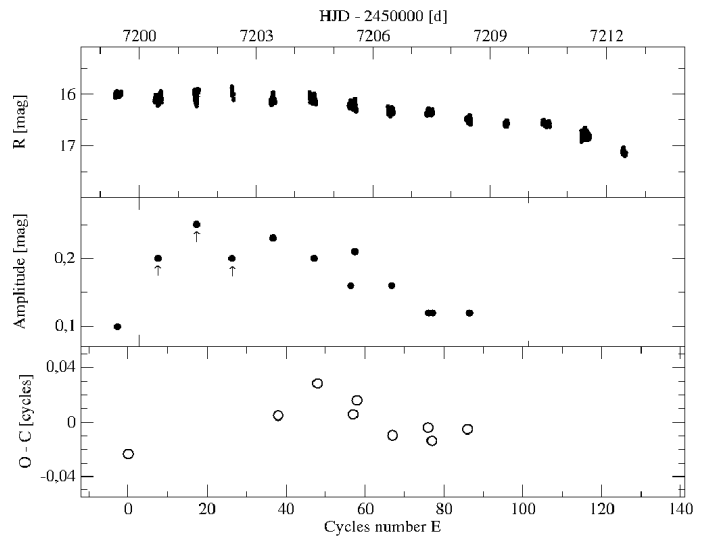}
    \caption{The light curves of MN Dra during the June-July 2015 superoutburst are shown in the top panel. The evolution of the amplitude of superhumps is presented in the middle panel. In three cases of scattered and/or incomplete data, the arrow marks a rough estimation of the amplitude which is 'larger than' the value indicated by a corresponding black circle.  The $O-C$ diagram for the superhumps maxima is in the bottom panel.}
    \label{fig:MNDra_OC_LC2_superout}
\end{figure}

Fig. \ref{fig:MNDra_OC_LC1_superout} and \ref{fig:MNDra_OC_LC2_superout} show light curves of MN Dra during its superoutbursts (top panels), evolution of amplitudes of humps (middle panels) and the $O-C$ diagrams for detected maxima (bottom panels). 

\begin{figure}
	\centering
	\includegraphics[width=\columnwidth]{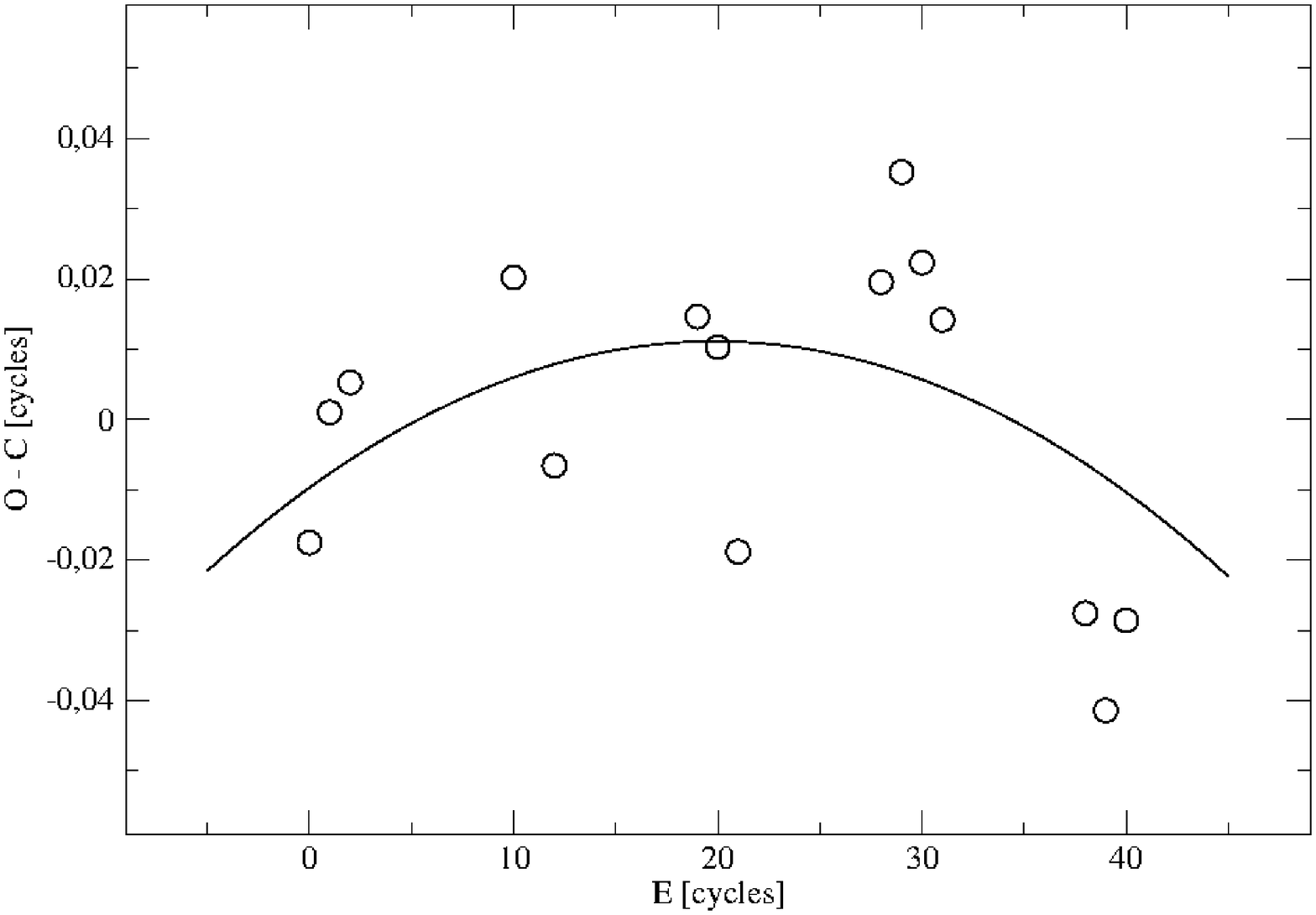}
    \caption{The $O-C$ diagrams for the September 2013 superoutburst in MN Dra with the second-order polynomial fit (black curve).}
    \label{fig:MNDra_OC_Sept2013}
\end{figure}

\begin{figure}
	\centering
	\includegraphics[width=\columnwidth]{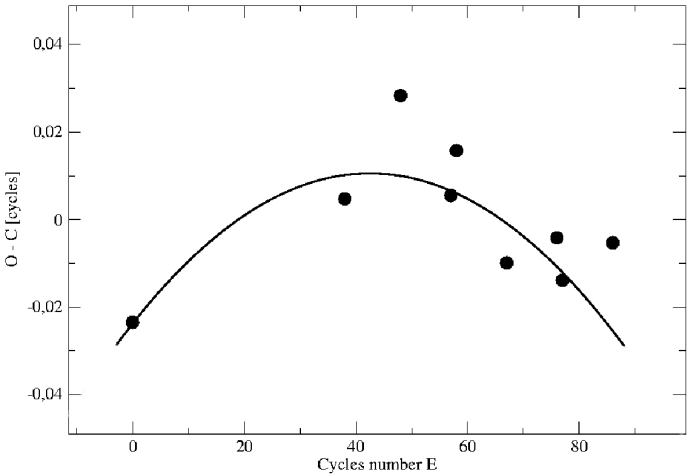}
    \caption{The $O-C$ diagrams for the June - July 2015 superoutburst in MN Dra with the second-order polynomial fit (black curve).}
    \label{fig:MNDra_OC_JuneJuly2015}
\end{figure}

We display the $O-C$ values corresponding to the both analysed superoutbursts from the September 2013 and the June - July 2015 in Fig. \ref{fig:MNDra_OC_Sept2013} and \ref{fig:MNDra_OC_JuneJuly2015}, respectively. Atmospheric conditions allowed us to obtain better coverage of the first part of the September 2013 superoutburst and the second part of the June - July 2015 superoutburst. Due to this fact in the September 2013 data most of the moments of maxima were detected between $0$ and $40$ cycles, and for the June - July 2015 set between $40$ and $90$ cycles.

In Table \ref{tab:MNDra_OC} we present superhump periods of MN Dra calculated until now. As aforementioned, we derived the second-order polynomial fit (black curves in Fig.  \ref{fig:MNDra_OC_Sept2013} and \ref{fig:MNDra_OC_JuneJuly2015}). However, in both cases the error of square factor $E^2$ in Eq. \ref{eq:MNDra_OCsq1} and \ref{eq:MNDra_OCsq2} is substantial enough to postulate that there is a poor agreement between the second-order polynomial fit and the $O-C$ values of the moments of maxima. Therefore, we cannot confirm the values of increasing or decreasing trends of the superhump period postulated by \cite{Nogami2003,Pavlenko2010,Samsonov2010}. Also, we question the value given by \cite{Kato2014b} due to its enormous error.

\begin{table}
	\centering
	\caption[Times of superhumps maxima of MN Draconis during its 2013 September superoutburst]{Times of superhump maxima of MN Dra during its 2013 September superoutburst.}
	\label{tab:OC_Sep2013}
	\begin{tabular}{ccccc} 
		\hline
		  \noalign{\smallskip}
		Cycle no. & Times of max. & Error & $O-C$  & Amplitude\\
		$E$           & HJD-2450000      &       & [cycles]& $A$ [mag]  \\
		  \noalign{\smallskip}		
		\hline
		  \noalign{\smallskip}
		0  & 6539.3258 & 0.001 & -0.0170 & 0.29 \\
		  \noalign{\smallskip}
		1  & 6539.4327 & 0.001 & 0.0015  & 0.34 \\
		  \noalign{\smallskip}
		2  & 6539.5381 & 0.002 & 0.0057  & 0.36 \\
		  \noalign{\smallskip}
		10 & 6540.3793 & 0.003 & 0.0203  & 0.39 \\
		  \noalign{\smallskip}
		12 & 6540.5864 & 0.001 & -0.0066 & 0.34 \\
		  \noalign{\smallskip}
		19 & 6541.3233 & 0.001 & 0.0143  & 0.25 \\
		  \noalign{\smallskip}
		20 & 6541.4278 & 0.002 & 0.0099  & 0.21 \\
		  \noalign{\smallskip}
		21 & 6541.5297 & 0.002 & -0.0192 & 0.22 \\
		  \noalign{\smallskip}
		28 & 6542.2684 & 0.002 & 0.0188  & 0.20 \\
		  \noalign{\smallskip} 
		29 & 6542.3750 & 0.001 & 0.0344  & 0.21 \\
		  \noalign{\smallskip}
		30 & 6542.4786 & 0.002 & 0.0214  & 0.22 \\
		  \noalign{\smallskip}
		31 & 6542.5827 & 0.002 & 0.0133  & 0.26 \\
		  \noalign{\smallskip}
		38 & 6543.3130 & 0.001 & -0.0288 & 0.25 \\
		  \noalign{\smallskip} 
		39 & 6543.4165 & 0.002 & -0.0427 & 0.25 \\
		  \noalign{\smallskip} 
		40 & 6543.5228 & 0.003 & -0.0299 & 0.21 \\
		  \noalign{\smallskip}
		48 & 6544.3665 & 0.002 & 0.0085  & 0.31 \\
		  \noalign{\smallskip}				
		\hline
	\end{tabular}
\end{table}

\begin{table}
	\centering
	\caption[Times of superhumps maxima of MN Draconis during its 2015 June-July superoutburst]{Times of superhumps maxima of MN Dra during its 2015 June - July superoutburst.}
	\label{tab:OC_Superoutburst_Jun2015}
	\begin{tabular}{ccccc} 
		\hline
		  \noalign{\smallskip}
		Cycle no. & Times of max. & Error & $O-C$ & Amplitude \\
		$E$           & HJD-2450000      &       & [cycles] & $A$ [mag] \\
		  \noalign{\smallskip}		
		\hline
		  \noalign{\smallskip}
		0  & 7199.4359 & 0.002 & -0.0232 & 0.10 \\
		  \noalign{\smallskip}
		38 & 7203.4335 & 0.001 & 0.0050 & 0.23 \\
		  \noalign{\smallskip}
		48 & 7204.4872 & 0.002 & 0.0286 & 0.20 \\
		  \noalign{\smallskip}
		57 & 7205.4309 & 0.001 & 0.0058 & 0.16 \\
		  \noalign{\smallskip}
		58 & 7205.5371 & 0.003 & 0.0160 & 0.21 \\
		  \noalign{\smallskip}
		67 & 7206.4805 & 0.001 & -0.0097 & 0.16 \\
		  \noalign{\smallskip}
		76 & 7207.4272 & 0.003 & -0.0039 & 0.12 \\
		  \noalign{\smallskip}
		77 & 7207.5313 & 0.002 & -0.0136 & 0.12 \\
		  \noalign{\smallskip}
	    86 & 7208.4783 & 0.005 & -0.0051 & 0.12 \\
	      \noalign{\smallskip}
		\hline
	\end{tabular}
\end{table}

\begin{table*}
	\centering
			\vspace{1cm}
	\caption[Values of superhump periods of MN Draconis from 2002 until 2015]{Values of superhump periods of MN Dra form 2002 until 2015}
	\label{tab:MNDra_OC}
	\begin{tabular}{llll} 
		\hline
		\noalign{\smallskip}
		  & Periods  & Rate of change \textit{\.{P}} & Author\\
		  & [days] &  & \\ 
		 \noalign{\smallskip}
		\hline
		\noalign{\smallskip}
		
	 October 2002 & $0.104885(93)$ & increasing & \cite{Nogami2003}\\
		\noalign{\smallskip}
		 December 2002 & $0.10623(16)$ & decreasing ($-1.7(2) \times 10^{-3}$) & \cite{Nogami2003}\\
		\noalign{\smallskip}
		 May 2009 & $0.105416(44)$ & decreasing ($-24.5 \times 10^{-5}$) & \cite{Pavlenko2010}\\
		\noalign{\smallskip}
		 July 2009 & $0.105416$ & decreasing ($-3.2 \times 10^{-4}$) & \cite{Samsonov2010} \\		
		\noalign{\smallskip}
		 September 2009 & $0.105416$ & decreasing ($-8.3 \times 10^{-4}$) & \cite{Samsonov2010} \\		
		\noalign{\smallskip}
		 July - August 2012 & $0.105299(61)$ & no data available & \cite{Kato2014b}\\
		\noalign{\smallskip}
		 September 2013  &  $0.10496(2)$ & no changes & This work \\
		\noalign{\smallskip}
		 November 2013 & $0.105040(66)$ & decreasing ($-14.8(9.5) \times 10^{-5} $) & \cite{Kato2014b}\\
		\noalign{\smallskip}
		June - July 2015&  $0.10512(3)$ & no changes & This work\\
		\noalign{\smallskip}		
		\hline
	\end{tabular}
\end{table*}

\subsubsection{Normal outbursts}

We performed the $O-C$ analysis for two sets of observations covering normal outbursts that occurred in June 2015 and in August 2015. 

During each normal outburst in MN Dra, we collected data during two subsequent nights, and identified 4 moments of maxima. All determined peaks of maxima, together with their errors, cycle numbers $E$ and the $O-C$ values are listed in Table \ref{tab:MNDra_OC_JunANDAug2015}.

From a linear fit, we calculated the following ephemerides:

\begin{equation}
{\rm HJD_{\rm max}} = 2457178.428(3) + 0.0955(3) \times E.
\label{eq:MNDra_OC_NO1}
\end{equation}

\begin{equation}
{\rm HJD_{\rm max}} = 2457248.418(2) + 0.0953(3) \times E.
\label{eq:MNDra_OC_NO2}
\end{equation} 

\noindent
and these values correspond to the negative superhump periods of $P_{nsh4}=0.0955(3)$ days ($137.52\pm0.4$ min) and $P_{nsh5}=0.0953(3)$ days ($137.23\pm0.4$ min) for the June 2015 and the August 2015 normal outbursts in MN Dra, respectively. 

\begin{figure}
	\centering
	\includegraphics[width=\columnwidth]{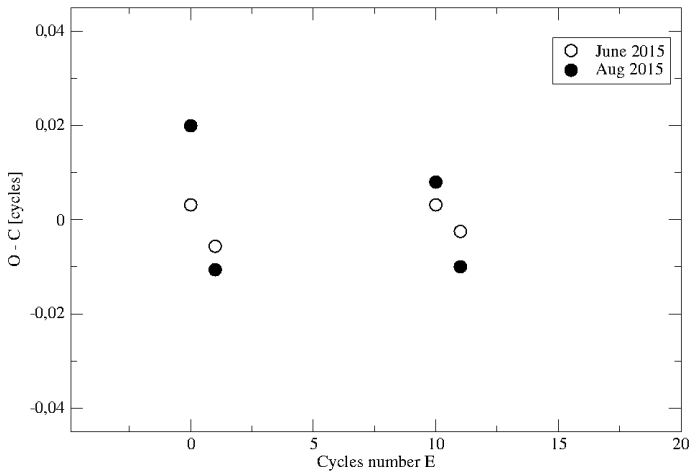}
    \caption{The $O-C$ diagram for the June 2015 (open circles) and the August 2015 (black circles) normal outbursts in MN Dra.}
   \label{fig:MNDra_OC_Outbursts}
\end{figure}

Fig. \ref{fig:MNDra_OC_Outbursts} displays the $O-C$ values corresponding to the ephemerides given by Eq. \ref{eq:MNDra_OC_NO1} and \ref{eq:MNDra_OC_NO2}.

\begin{table}
	\centering
	\caption[Times of negative superhump maxima of MN Draconis during its normal outbursts in June and August 2015]{Times of negative superhump maxima of MN Dra during its normal outbursts in June and August 2015.}
	\label{tab:MNDra_OC_JunANDAug2015}
	\begin{tabular}{cccr} 
		\hline
		  \noalign{\smallskip}
		Cycle no. & Times of max. & Error & $O-C$ \\
		$E$           & HJD-2450000      &       & [cycles] \\
		  \noalign{\smallskip}		
		\hline
		  \noalign{\smallskip}
		0  &  7178.4283 & 0.003 &  0.0031  \\
		  \noalign{\smallskip}
        1  &  7178.5230 & 0.004 & -0.0057  \\
          \noalign{\smallskip}
        10 &  7179.3837 & 0.003 &  0.0031  \\
          \noalign{\smallskip}
        11 &  7179.4787 & 0.004 & -0.0025  \\
          \noalign{\smallskip}
		\hline
		  \noalign{\smallskip}
		0  & 7248.4200 & 0.003 &  0.0199 \\
		  \noalign{\smallskip}
        1  & 7248.5124 & 0.002 & -0.0106 \\
          \noalign{\smallskip}
        10 & 7249.3720 & 0.003 &  0.0080\\
          \noalign{\smallskip}
        11 & 7249.4656 & 0.004 & -0.0100 \\
          \noalign{\smallskip}
		\hline
	\end{tabular}
\end{table}

\subsubsection{Quiescence}

Furthermore, the $O-C$ diagrams were constructed for two campaigns conducted in October 2009 and in October 2015 while MN Dra was in quiescent state. 

Once again, in Table \ref{tab:MNDRa_OC_Q} we present the HJD times of detected moments of maxima, errors, number of cycles $E$ and the $O-C$ computed according to the linear ephemerides presented below:

\begin{equation}
{\rm HJD_{\rm max}} = 2455117.314(2) +  0.09643(3) \times E.
\label{eq:MNDra_OC_Q1}
\end{equation}

\begin{equation}
{\rm HJD_{\rm max}} = 2457308.6436(9) + 0.09541(6) \times E.
\label{eq:MNDra_OC_Q2}
\end{equation} 

\noindent
and associated to the negative superhump periods of $P_{nsh6}=0.09643(3)$ days ($138.8592\pm0.04$ min) and $P_{nsh7}=0.09541(6)$ days ($137.3904\pm0.09$ min) for the October 2009 and the October 2015 light curves of MN Dra in quiescence, respectively. 

\begin{table}
	\centering
	\caption[Times of negative superhump maxima of MN Draconis during quiescence in October 2009 and October 2015]{Times of negative superhump maxima of MN Dra during quiescence in October 2009 and October 2015.}
	\label{tab:MNDRa_OC_Q}
	\begin{tabular}{cccr} 
		\hline
		  \noalign{\smallskip}
		Cycle no. & Times of max. & Error & $O-C$ \\
		$E$           & HJD-2450000      &       & [cycles]  \\
		  \noalign{\smallskip}		
		\hline
		  \noalign{\smallskip}
		0  & 5117.3165 & 0.003 &  0.0023  \\
		  \noalign{\smallskip}
        20 & 5119.2420 & 0.002 & -0.0096  \\
          \noalign{\smallskip}
        21 & 5119.3384 & 0.003 & -0.0099  \\
          \noalign{\smallskip}
        73 & 5124.3542 & 0.002 &  0.0050  \\
          \noalign{\smallskip}
        93 & 5126.2822 & 0.003 & -0.0012  \\
          \noalign{\smallskip}
		\hline
		  \noalign{\smallskip}
		0  & 7308.6439 & 0.001 &  0.0031  \\
		  \noalign{\smallskip}
        1  & 7308.7369 & 0.003 & -0.0221  \\
          \noalign{\smallskip}
        11 & 7309.6938 & 0.004 & -0.0094  \\
          \noalign{\smallskip}
        21 & 7310.6478 & 0.001 &  0.0072  \\
          \noalign{\smallskip}
        22 & 7310.7408 & 0.002 & -0.0191  \\
          \noalign{\smallskip}
		\hline
	\end{tabular}
\end{table}

\begin{figure}
	\centering
	\includegraphics[width=\columnwidth]{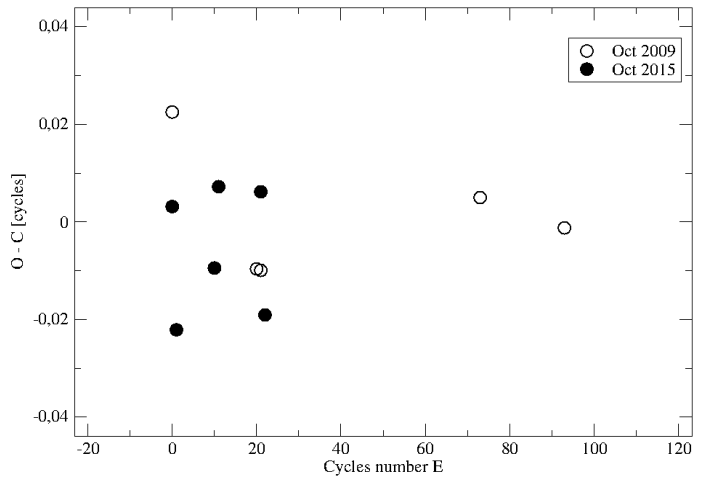}
    \caption{The $O-C$ diagram for the October 2009 (open circles) and the October 2015 (black circles) data sets of MN Dra in quiescence.}
   \label{fig:MNDra_OC_Outbursts_Quiescence}
\end{figure}

In Table \ref{tab:MNDra_OC2} we present detected periods, from the $O-C$ analyses, in the light curves of MN Dra during its  normal outbursts and in quiescence. Differences between the values of $P_{nsh}$ presented in Table \ref{tab:MNDra_OC2} excess $3\sigma$. That indicates that the negative superhump period of MN Dra is not stable. The mechanism responsible for negative superhumps is probably a classical 'regression' of the tilted disk. Therefore, in the case of MN Dra, the period of the 'regression' of the accretion disk is also changing.

The $O-C$ values corresponding to the ephemerides given by Eq. \ref{eq:MNDra_OC_Q1} and \ref{eq:MNDra_OC_Q2} are presented in Fig.,\ref{fig:MNDra_OC_Outbursts_Quiescence}.

\begin{table*}
	\centering
	\caption[Values of negative superhump periods of MN Draconis from 2002 until 2015]{Values of negative superhump periods of MN Dra from 2002 until 2015.}
	\label{tab:MNDra_OC2}
	\begin{tabular}{lll} 
		\hline
		\noalign{\smallskip}
		  & Periods [days] & Author\\
		 \noalign{\smallskip}
		\hline
 August 2002 - February 2003& 0.10424(3) & \cite{Nogami2003} \\	 	
		\noalign{\smallskip}
		October 2009   &  $0.09643(3)$ &
		 This work\\
		\noalign{\smallskip}
		 May - June 2009 & 0.09598(2) & \cite{Pavlenko2010} \\
		\noalign{\smallskip}
		 August - November 2009 & 0.095952(4) & \cite{Samsonov2010} \\
		\noalign{\smallskip}	
		 June 2015   &  $0.0955(3)$ & This work\\
		\noalign{\smallskip}		
		 August 2015   &  $0.0953(3)$ & This work\\   
		\noalign{\smallskip}	
		October 2015  &  $0.09541(6)$ & This work\\
		\noalign{\smallskip}			
		\hline
	\end{tabular}
\end{table*}

\section{Discussion}
\label{Discussion}

\subsection{Supercycle length}

The supercycle lengths of several very active DN below the period gap have been already analysed by \cite{Otulakowska2013a} and \cite{Otulakowska2013b}. They showed observational evidence that the supercycle lengths of investigated systems have been increasing as a result of their mean mass-transfer rates which have been constantly decreasing. Therefore, they checked out this phenomenon in the context of the future evolution of DN. With the assumption that the supercycle length will be increasing in the same way in the future, they estimated the time-scale of next steps of evolution of ER UMa objects to become SU UMa and WZ Sge-type systems (see Table 3 in  \citealt{Otulakowska2013b}). These predictions are in agreement with results presented by \cite{Patterson2013} concerning the BK Lyn evolution. According to \cite{Patterson2013}, BK Lyn, as being a member of ER UMa group nowadays, is in a transient stage of evolution, preceded by the classical nova and nova-like variable phases. If this hypothesis is true for all active SU UMa systems, that means all these objects are survivors of classical nova eruptions which have been fading ever since.

The changes in supercycle length of MN Dra are the same as the behaviour of several cases of DN below the period gap presented by \cite{Otulakowska2013a} and \cite{Otulakowska2013b}. Therefore, we discovered the first period gap object in which the occurrence of superoutbursts has been constantly decreasing within last decades.

\subsection{Orbital period determination}

We used the latest version of the  Stolz and Schoembs relation \cite{Otulakowska2016}:

\begin{equation}
{\rm log }\: \varepsilon = 1.97 (0.10) \times {\rm log}\: P_{orb}\: {\rm [d]} + 0.73(0.11),
\end{equation}

\noindent
and the following formula defining the superhump period excess/deficit as

\begin{equation}
\varepsilon = \frac{P_{sh}-P_{orb}}{P_{orb}},
\end{equation}

\noindent
to estimate the orbital period $P_{orb}= 0.0994(1)$ days ($143.126 \pm 0.144$ min) and the period excess $\varepsilon=5.7\% \pm 0.1\%$ and deficit $\varepsilon_-=2.5\% \pm 0.6\%$ and their ratio $\phi=-0.44(11)$ for MN Dra. 

\begin{figure}
	\vspace{0.5cm}
	\centering
	\includegraphics[width=\columnwidth]{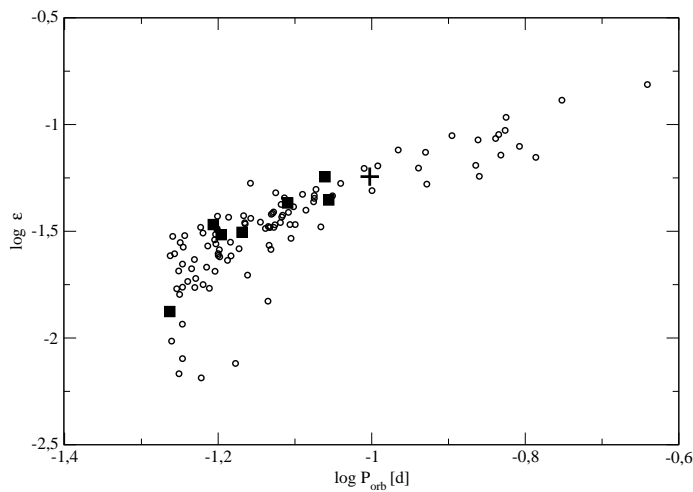}
    \caption{The relation between the period excess and the orbital period for DN. Small, open circles represent known DN \citep{Olech2011}. Black squares correspond to the position of active systems for which the supercycle lengths were calculated \citep{Otulakowska2013b}. The position of MN Dra is marked with black cross.}
   \label{fig:MNDra_Period_excess}
\end{figure}

Knowing $\varepsilon$ and $P_{orb}$, we were able to check the evolutionary status of DN, since the mass ratio decreases with time due to the mass-loss from the secondary. In Fig. \ref{fig:MNDra_Period_excess} small, open circles represent known DN (from \citealt{Olech2011}) and black squares correspond to systems with known supercycle lengths (from \citealt{Otulakowska2013b}). The position of MN Dra is marked with black cross.

\cite{Retter2002} suggested that there is a correlation between $\phi$ and $P_{orb}$ for CVs which exhibit both positive and negative superhumps and \cite{Olech2009} presented the following empirical formula for this dependency:

\begin{equation}
\phi = 0.318(6) \times {\rm log}\:P_{orb}\: {\rm [d]} - 0.161(10).
\end{equation}

\begin{figure}
	\vspace{0.5cm}
	\centering
	\includegraphics[width=\columnwidth]{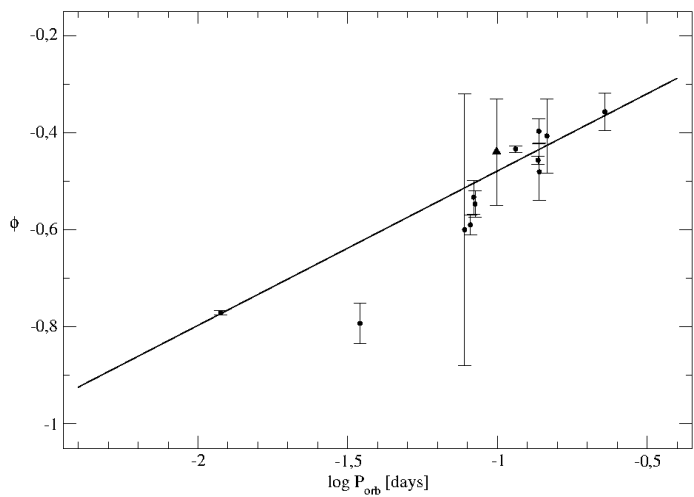}
    \caption{The relation between the ratio between period deficit and excess and orbital period of different types of CVs. The position of MN Dra is marked by black triangle. Figure taken from \citep{Olech2009}.}
   \label{fig:MNDra_Period_excess2}
\end{figure}

\noindent
We can confirm their hypothesis for the case of MN Dra. Fig. \ref{fig:MNDra_Period_excess2} presents the relation between the ratio $\phi$ and orbital period for several CVs including MN Dra.

Next, by employing an empirical formula for the relation between the period excess and the mass ratio of the binary $q=M_2/M_1$ \citep{Patterson1998}:

\begin{equation}
\varepsilon=\frac{0.23q}{1+0.27q},
\end{equation}

\noindent
we derived the mass ratio for MN Dra as equal to $q \approx 0.26$. It is worth to mention that \cite{Kato2014b} obtained similar result of $q=0.258$ based on the orbital period calculated by \cite{Pavlenko2010}.

\subsection{Superhumps}

\subsubsection{Active period-gap DN}

In CVs with long orbital periods, magnetic braking is the dominant mechanism for angular momentum loss. It is thought that around $P_{orb} \backsimeq 3$ h the secondary becomes fully convective \citep{Verbunt1981} and causes the termination of magnetic braking. In response, the secondary contracts and detaches from its Roche lobe. At this point, the cataclysmic system is a detached binary of low luminosity. Due to the continuous loss of orbital angular momentum, the orbit decays and this results in the re-establishment of contact at the period of $\sim 2$ hours. Hence, the mass transfer is resumed and both its loss as well as its rate are driven by gravitational radiation. The system re-emerges as an active CV at the bottom of the period gap. \cite{Knigge2006} revisited the estimation of the period gap and presented its new localisation at $3.18(0.04) \leq P_{orb} \leq 2.15(0.03) $ hours. 

Even though the explanation outlined above seems satisfactory for the significant dearth of CVs in the period range $2 \lesssim P_{orb} \lesssim 3$ h \citep{Gansicke2009}, there are still many missing pieces in this puzzle. SU UMa systems located in the period gap are expected to have neither significant magnetic breaking nor significant angular momentum loss from gravitational radiation. Therefore these objects should be characterized by low activity. Hence, among the challenges open to interpretation, are the recent discoveries of active DN with orbital periods within period gap, such as TU Men \citep{Stolz1981}, SDSS J162520.29+120308.8 \citep{Olech2011}, OGLE-BLG-DN-001 \citep{Poleski2011}, CzeV404 \citep{Bakowska2014} or NY Ser \citep{Pavlenko2014}. Taking into account that MN Dra has an orbital period within the period gap range ($P_{orb} \sim 2.38$ hours) and shows rapid superoutburst activity ($P_{sc} \sim 74$ days), this DN is the latest observational evidence contradicting the existing  theory of evolution of CVs.

\subsubsection{Positive superhumps}

To check whether the rate of \textit{\.{P}} changes from epoch-to-epoch significantly, we investigated all published \textit{O-C} diagrams of positive superhumps in MN Dra. Our conclusion is that during most of the superoutbursts MN Dra shows smooth changes of superhump period and rather small/standard negative superhump period derivatives. Frequently, the $O-C$ diagrams look rather 'flat' during the whole superoutburst (see Fig. 6 in \citealt{Nogami2003} and Fig. 2 in \citealt{Samsonov2010}). 

\cite{Kato2009, Kato2010, Kato2012, Kato2013, Kato2014a, Kato2014b, Kato2016a} published a comprehensive survey of period variations of superhumps in SU UMa-type DN. According to them, the evolution of the superhump period usually can be divided into three separate parts: the first stage (A) corresponds to an early superhumps with a stable and longer period, the middle stage (B) refers to positive superhump period derivative and the final stage (C) characterizes with a shortened but again stable superhump period. According to \cite{Kato2009}, in the well-observed DN, the transition between stages A and B, and stages B and C are abrupt and discontinuous. There are examples of SU UMa period-gap objects following this scenario, e.g. V1006 Cyg \citep{Kato2016c} and SDSS J162520.29+120308.8 \citep{Olech2011}. However, \cite{Olech2003} showed that these transitions can be also  smooth changes. In case of MN Dra, \cite{Kato2014b} presented the transition between stages A and B for several superoutbursts (see Fig. 19 in \citealt{Kato2014b}) and  concluded that the superhump period of MN Dra is characterized by a large negative \textit{\.{P}}. Nonetheless, in the $O-C$ diagrams constructed on our data sets we did not detect stage A of MN Dra during any of its superoutbursts. Moreover, it seems that lack of significant or abrupt changes in superhump period in MN Dra is not peculiar. There are cases of DN which display rather small changes in superhump derivatives and the shapes of their $O-C$ diagrams are rather 'flat', e.g. RZ LMi (Fig. 2 in \citealt{Kato2016b}), V452 Cas (Fig.6 in \citealt{Kato2014b}), or GZ Cnc (Fig.11 in \citealt{Kato2014b}). Hence, the observed behaviour of MN Dra is in agreement with theory which interprets negative superhumps period derivatives in the $O-C$ diagrams as a result of the disk shrinkage during the superoutburst and thus lengthening its precession rate \citep{Lubow1991,Patterson1998}.

\subsubsection{Negative superhumps}

CVs with negative superhumps' manifestation are not frequently detected. Up-till-now, we know roughly a dozen of confirmed cases (see Table 2 in \citealt{Montgomery2009}, Table 5 in \citealt{Armstrong2013}). Unlike positive superhump theory, no consensus about the source of negative superhump has yet been established. Possible mechanisms responsible for negative superhumps and challenges open for interpretation can be found in \cite{Montgomery2009}. 

In case of MN Dra, negative superhumps were investigated during normal outbursts by \cite{Samsonov2010} and \cite{Pavlenko2010}. They noticed that maxima of negative superhumps varied cyclically in correlations with normal outbursts (Fig. 5 and Fig. 6 in \citealt{Samsonov2010}, Fig. 5 in \citealt{Pavlenko2010}). Based on our results, we cannot confirm this correlation. However, it must be noted that our sets of data, gathered during normal outbursts in MN Dra, are scattered and insufficient for any conclusive remarks.

According to \cite{Patterson1997}, negative superhumps are associated with retrograde precession and are not strictly periodic. For example, these signals are present during several months of observations, they cannot be detected a year later, and finally they reappear several years after that. In this context, a disk can remain stably tilted for some time, later on the disk re-aligns with the orbital plane and maintains this position for another period of time, then it returns to its previous tilted position. Nonetheless, the problem of how a disk initially tilts remains unsolved. We think that MN Dra follows that scenario. We did not detect any changes in the negative superhump periods on the short time-scales of days and months during normal outbursts and during quiescence. That is why our $O-C$ diagrams were rather 'flat'. However, we noticed the differences between the values of $P_{nsh}$ from 2009 and 2015 which excess $3\sigma$.  That shows that the negative superhump period of MN Dra is not stable on the longer time-scales of years. Hence, it is probable that the mechanism responsible for negative superhumps is a classical 'regression' of the tilted disk.

It is worth noting that our observational results of the period excess, period deficit and mass ratio of MN Dra are in agreement with the recent smoothed particle hydrodynamics simulations performed by \cite{Thomas2015} (check Fig. 9). Therefore, MN Dra is an excellent object to test thoroughly the hypothesis of the tilted-disk geometry as the source of negative superhumps, but most importantly, to test the theory which invokes white-dwarf magnetism to break the azimuthal symmetry and permits to create disk tilt postulated by \cite{Thomas2015}.

\section{Conclusions}
\label{Conclusions}

To conclude, we present a summary of our world-wide observational campaign of MN Dra.

\begin{itemize}

\item MN Dra was observed during three campaigns in 2009, 2013, and 2015. We detected MN Dra in a quiescent state during 6 nights of observations in October 2009. At the time of our second campaign, between June and September 2013, the star was monitored during two superoutbursts interspersed with one normal outburst. In our most recent campaign, from June to December 2015, two superoutbursts and three normal outbursts in MN Dra were recorded. The average amplitude of brightness during superoutbursts was $A_s \approx  2.9$ mag. We detected clear positive superhumps during superoutburts. Also, negative superhumps were observed in this DN during its normal outbursts and in quiescence. The duration of a superoutburst was on average $24\pm 1$ days.

\item The supercycle length, with a value of $P_{sc}=74\pm0.5$ days, was derived for the two subsequent superoutbursts observed in June - July 2015 and in September 2015. Based on our data set and observations presented by \cite{Nogami2003} and \cite{Samsonov2010}, the supercycle length has been increasing during the last twelve years with a rate of $\textit{\.{P}}= 3.3 \times 10^{-3}$. Additionally, the occurrence of superoutbursts observed in MN Dra follows the scenario presented in \cite{Otulakowska2013a} and \cite{Otulakowska2013b} for very active ER UMa stars. Also, MN Dra is the first discovered SU UMa system in the period gap with increasing supercycle length, and therefore it is crucial for our understanding of the future evolution of DN.

\item We conducted the $O-C$ analysis for the moments of maxima detected in the September 2013 and the June - July 2015 superoutbursts in MN Dra, and the superhump period was $P_{sh4}=0.10496(2)$ days ($151.1424\pm 0.03$ min) and $P_{sh5}=0.10512(3)$ days ($151.3728\pm 0.04$ min), respectively. Also, the second-order polynomial fit was calculated. Nonetheless, due to the poor agreement between the fit and the $O-C$ values of the moments of maxima, we cannot confirm the changes in trend of the superhump period postulated by \cite{Nogami2003}, \cite{Samsonov2010} or \cite{Pavlenko2010}.

\item To investigate the negative humps in MN Dra, we used our best quality observations obtained in October 2009 and in October 2015 on 1.3 meter telescopes located in Greece and in the U.S.A., respectively. The negative superhump period was obtained with a value of $P_{nsh6}=0.09643(3)$ days ($138.8592\pm0.04$ min) and $P_{nsh7}=0.09541(6)$ days ($137.3904\pm0.09$ min) for the 2009 and the 2015 data, respectively. 

\item We derived the period excess and the period deficit $\varepsilon= 5.7\% \pm 0.1\%$ and $\varepsilon_-= 2.5\% \pm 0.6\%$, respectively. Also, the orbital period with the value of $P_{orb}=0.0994(1)$ days ($143.126 \pm 0.144$ min) was obtained. On the diagram $P_{orb}$ versus $\varepsilon$, MN Dra is located in a range of the period gap objects. 

\item We obtained the mass ratio with the value of $q \approx 0.26 $ which is in accordance with one of the values, $q=0.258$, obtained by \cite{Kato2014b} .

\end{itemize}

MN Dra is another example of an active DN located in the period gap \citep{Bakowska2014}. It is worth noting that this star is not only a challenge for existing models of the superhump and superoutburst mechanisms, but also it presents other intriguing behaviours, in particular the increasing supercycle length. Moreover, we know only several DN which exhibit positive and negative superhumps (see \citealt{Retter2002}, \citealt{Olech2009}) and there are the only two known cases of period gap SU UMa objects showing the negative superhumps \citep{Pavlenko2016}. Hence, MN Dra is a perfect object for further photometric observations, i.e. a good light curve coverage in normal outbursts would allow to determine normal cycle length and check its stability. To conclude, we presented a new set of information about MN Dra. Also, we updated available basic statistics of the system, i.e. the orbital period where we used different than \cite{Pavlenko2010} indirect methods for its determination.
 
We emphasize the lack of spectroscopic analysis of this object which could provide many answers regarding this kind of very active, period gap systems. However, a very low (below 18 mag) brightness in quiescence of MN Dra, located on the northern sky, poses a challenge for spectroscopic observations.

\section*{Acknowledgements}

This work is partially based on observations obtained at the MDM Observatory, operated by Dartmouth College, Columbia University, Ohio State University, Ohio University, and the University of Michigan. KB wants to thank K.Z. Stanek and E. Galayda for the time allowance and the technical support in MDM Observatory. We are indebted to an anonymous reviewer of an earlier version of  this paper for providing insightful comments and providing directions for additional work which has resulted in this paper. The project was supported by the Polish National Science Center grants
awarded by decisions DEC-2012/07/N/ST9/04172 and DEC-2015/16/T/ST9/00174 for KB.


\bibliographystyle{aa}

\end{document}